\documentclass[a4paper,11pt]{article}

\usepackage{latexsym}
\usepackage{amsmath}
\usepackage{amssymb}
\usepackage{bbm}
\usepackage{cite}
\usepackage[dvips]{graphicx} 
\usepackage{mathrsfs}

\newcommand{\ve}{\varepsilon}
\newcommand{\halb}{\frac{1}{2}}
\newcommand{\hal}{{\textstyle\frac{1}{2}}}

\newcommand{\ca}[1]{{\cal{#1}}}

\newcommand{\unit}{\mathbbm{1}} 
\newcommand{\fr}[2]{{\textstyle{\frac{#1}{#2}}}}
\newcommand{\vf}{\varphi}
\newcommand{\slashed}[1]{#1\! \! \!\! /}
\newcommand{\slasha}[1]{#1\! \! \! /}
\newcommand{\mD}{\mathscr{D}}

\newcommand{\bra}[1]{\left\langle#1\right|}
\newcommand{\ket}[1]{\left|#1\right\rangle}

\begin{document}

%%%%%%%%%%%%%%%%%%%%%%%%Titlepage%%%%%%%%%%%%%%%%%%%%%%%%%%%%%%%%%%%%%%%%%%%%%%%%
\begin{titlepage}
    \renewcommand{\thefootnote}{\alph{footnote}}
    ~\vspace{-2cm}
    \begin{flushright} 
      hep-th/0502158\\
      ITP-UH-03/05
    \end{flushright}  
    \vfil
  \centerline{\Large\bf $D0$-$D4$ brane tachyon condensation to a BPS state} 
    \medskip
  \centerline{\Large\bf and its excitation spectrum in}
    \medskip
  \centerline{\Large\bf   noncommutative super Yang-Mills theory}
    \medskip
\begin{center}
 \vfil 
 {\large\sc
    Robert Wimmer$^*$\footnotetext{\footnotesize${}^*$\tt wimmer@itp.uni-hannover.de}
  }\\
\end{center}  \medskip \smallskip \qquad \qquad 
\begin{center}
 {\sl 
  Institut f\"ur Theoretische Physik, Universit\"at Hannover,\\ 
  Appelstr.~2, D-30167 Hannover, Germany\\ }
\end{center}
\vfil
\centerline{ABSTRACT}\vspace{.5cm}

We investigate the $D0$-$D4$-brane system for different $B$-field backgrounds 
including the small instanton singularity in noncommutative SYM theory. 
We discuss the excitation spectrum of the 
unstable state as well as for the BPS $D0$-$D4$ bound state. We compute the tachyon 
potential which reproduces the complete mass defect. The relevant degrees of freedom
are the massless $(4,4)$ strings. Both results are in contrast with existing string field
theory calculations. The excitation spectrum of the small instanton is found to be 
equal to the excitation spectrum of the fluxon solution on 
$\mathbb{R}_\theta^2\times\mathbb{R}$ which we trace back to $T$-duality. For the 
effective 
theory of the $(0,0)$ string excitations we obtain a BFSS matrix model. The number
of states in the instanton background changes significantly when the $B$-field 
becomes self-dual. This leads us to the proposal of the existence of a phase 
transition or cross over at self-dual $B$-field.

\end{titlepage}
%%%%%%%%%%%%%%%%%%%%%%%%%%%%%%%%%%%%%%%%%%%%%%%%%%%%%%%%%%%%%%%%%%%%%%%%%%%%%%%%

\newpage

\tableofcontents

\section{Introduction}

The striking correspondence between superstring theories with $D$-branes
and supersymmetric gauge theories led to a number of 
insights in the structure of supersymmetric gauge theories. 
This correspondence becomes even more interesting if a 
 constant $B$-field background is taken into account in  string theory. 
The resulting low energy description is a noncommutative gauge theory where 
the noncommutative structure of space is determined by the $B$-field background
 \cite{Seiberg:1999vs}. 
In addition to $\alpha'$, the $B$-field introduces a new scale into the theory
  and in some cases 
much more of the  structure of the string theory survives the low energy limit, where 
$\alpha'\rightarrow 0$.
This is  observed as a large number of nontrivial classical solutions 
in the gauge theory and also the excitation spectrum which survives this limit
 shows much more 
diversity than in the case without a $B$-field background.  
Hence there is some legitimate hope that from noncommutative field theory one 
learns something about string theory, using the somewhat simpler techniques of
(noncommutative) quantum field theory. 
In the generic case 
the resulting noncommutativity also acts as a regulator at small distances. 
For example the small instanton singularity in the moduli space of 
instantons is resolved for a generic  $B$-field background.

In this paper we consider supersymmetric $U(1)$ Yang-Mills theory defined on
$\mathbb{R}^4_\theta\times\mathbb{R}$ or $\mathbb{R}^4_\theta$. We investigate 
various (quantum) aspects of nontrivial solutions on $\mathbb{R}^4_\theta$, which are
solitonic states in the five dimensional case or (unstable) ``instantons''
in the latter case. The $D$-brane setup for such configurations is
a $D0$-$D4$-brane system where the rank four $B$-field has nontrivial 
components along the $D4$-brane.

The different aspects that we address for this system can be
summarised in a cycle of different states of this system. Initially, for
generic $B$-field the $D0$-$D4$ system is unstable and  described by a solitonic unstable 
solution of the Yang-Mills equations. We review the spectrum of this system
 taking nontrivial scalar field backgrounds into account. They are identified
with the distance of the $D0$-brane from the $D4$-brane in the transverse directions.
If the separation of the two branes becomes smaller than the size 
of the solitonic state the system becomes unstable 
and a tachyonic mode appears. 
Then the system condensates to a BPS $D0$-$D4$ bound state. 
We calculate the tachyon potential and identify
the massless $(4,4)$ strings as the relevant degrees of freedom in the
tachyon condensation process. The depth of the tachyon potential 
equals exactly the mass defect in forming the $D0$-$D4$ bound state in the
considered Seiberg-Witten limit. Both observations for the tachyon condensation are 
in contrast to string field theory computations \cite{David:2000um}
as we will discuss. 
The 
BPS $D0$-$D4$ bound state is described by a noncommutative instanton with a 
non-self-dual $B$-field background. We compute the exact excitation spectrum
for this state and show that it is quantitatively and qualitatively 
very different from the 
spectrum of the unstable state. In the next step we tune the 
$B$-field to become self-dual. The $D0$-$D4$ bound state instanton solution
smoothly approaches the small instanton which is described by a regular 
self-dual solution although the moduli space has a small instanton singularity
as in the commutative case.  
We discuss the various zero-mode moduli of the small instanton solution  
and  its excitation spectrum. Although the classical solutions
are continuously connected by this $B$-field limit  the spectra are not.
The number of states is significantly different. New scalar field zero-modes
show that the $D0$-brane deconfines  and can separate from 
the $D4$-brane. We also find that the effective theory for $D0$-$D0$ 
string excitations is described by a BFSS matrix model. We observe
that the spectrum of the small instanton coincides with the spectrum 
of the fluxon solution on $\mathbb{R}_\theta^2\times\mathbb{R}$ 
\cite{Gross:2000ph,Polychronakos:2000zm}
  and
argue, based also on the ADHM construction, that this is due to $T$-duality.
Because of the significant change in the number of states we propose the existence of
a phase transition or crossover in the noncommutative super Yang-Mills theory
when the $B$-field becomes self-dual and the $D0$-brane separates from the
$D4$-brane.  
In principle one can now reinitiate this circuit by turning on again an anti-self-dual 
part of the $B$-field and driving the system  into its unstable state. 

For technical reasons the paper is not organised in the circuit as described 
above but as follows: In section 2 we review some basic facts about 
noncommutative field theory to set up our  notation. 
 In section 3 we show how our formalism includes 
lower dimensional solitons. 
In section 4 we discuss 
the unstable state and the small instanton.
In section 5 we investigate the
instanton solution for non-self-dual $B$-field. In section 6 we discuss the 
tachyon condensation and compute the tachyon potential. In section 7
we draw our conclusions and present a proposal as well as a puzzle.

\section{Noncommutative field theory}
To set up our notation we briefly review some of the basic results of
noncommutative field theory. Of particular interest 
are noncommutative Yang-Mills
theories which arise as a certain zero slope limit $\alpha'\rightarrow 0$ of 
open string theory in a $B$-field background \cite{Seiberg:1999vs}. 

In this so called Seiberg-Witten limit the three point 
amplitudes of mass-less open strings on a stack of $N$ 
$Dp$-branes are reproduced by
the noncommutative $U(N)$ Yang-Mills action 
\begin{equation}
  \label{eq:ncYM1}
  \frac{1}{2 g_{YM}^2}\int\! dx^{p+1}G^{MP}G^{NQ}\ 
\textrm{tr}\{F_{MN}\star F_{PQ}\} +{\cal{O}}(\alpha')\ ,
\end{equation}
where the open string metric $G_{MN}$ in the Seiberg-Witten limit is
expressed through the closed string metric $g_{MN}$ and the $B$-field $B_{MN}$
in the following way:
\begin{equation}
  \label{eq:osm}
  G_{MN}\rightarrow -(2\pi\alpha')^2(Bg^{-1}B)_{MN}\ \ .
\end{equation}
We take the gauge fields to be anti-hermitian and the anti-hermitian 
generators of the $U(N)$ gauge group are normalised to 
$\textrm{tr}\{T^aT^b\}=-\halb \delta^{ab}$. 
%The open string metric we assume to be flat in the following, i.e. 
%\linebreak[4]{$G_{MN}=(-,+,\dots, +)$} 
%in Cartesian coordinates w.r.t. the open string metric. 
The field strength 
\begin{equation}
  \label{eq:FMN}
  F_{MN}=\partial_M A_N-\partial_N A_M+[A_M,A_N]_\star\
\end{equation}
is also defined w.r.t. the star-product ``$\star$'', which for constant
$\theta^{MN}$ is given by 
\begin{equation}
  \label{eq:star}
  (f\star g)(x):=e^{\frac{i}{2}\theta^{MN}
 \partial_M\partial'_N}f(x)g(x')|_{x'=x}\quad.
\end{equation}
In the Seiberg-Witten limit the noncommutativity parameter 
and the Yang-Mills coupling are related to the 
closed string coupling $g_s$ and the background 
$B$-field as follows:
\begin{equation}
  \label{eq:ymst}
  \theta^{MN}\rightarrow(B^{-1})^{MN}\quad ,\quad 
  \frac{1}{g_{YM}^2}\rightarrow
  \frac{\alpha'^{\frac{3-p}{2}}}{(2\pi)^{p-2} g_s\sqrt{det(2\pi\alpha'Bg^{-1})}}\ .
\end{equation}
%The Seiberg-Witten limit can be  characterised by $2\pi\alpha'B\gg g$, 
%with $g$ being the closed string metric. 
In matching instanton quantities
with $D$-brane quantities we will need these relations.
 
Noncommutative geometry is therefore defined through the algebra
${\cal{A}}_\theta$ of functions on a manifold 
with the multiplication given by 
a star-product.
Our case of interest are $D4$-branes with a non-vanishing
$B$-field along the four (spatial) dimensions of the branes. 
This corresponds to the noncommutative $\mathbb{R}^4_\theta$, which we 
parametrise by Cartesian coordinates $x^{\mu=1,\dots,4}$. The 
star-product (\ref{eq:star}) leads to the fundamental 
relation for the algebra ${\cal{A}}_\theta$, 
\begin{equation}
  \label{eq:sscom}
  [x^\mu,x^\nu]_\star = i\theta^{\mu\nu}\ .
\end{equation}

In \cite{Seiberg:1999vs} the string spectrum for the $D0$-$D4$ brane system
 was computed for a flat 
target space, i.e. with the closed string metric $g_{\mu\nu}=\delta_{\mu\nu}$. 
%The $B$-field was taken in its canonical form 
%$B_{\mu\nu}=(2\pi\alpha')^{-1}(b_1\ve,b_2\ve)$ with $\ve_{12}=1$. 
We will rather work
in Cartesian coordinates with respect to the open string metric  
(\ref{eq:osm}), i.e. $G_{\mu\nu}=\delta_{\mu\nu}$. 
This involves a $B$-field dependent coordinate transformation (see below).\\ 
    
\noindent
{\bf{Operator formalism.}} The algebra ${\cal{A}}_\theta$ can also be represented 
 in a different way. When  representing the generators $x^\mu$ by 
operators $\hat x^\mu$ acting on a Hilbert space ${\cal{H}}$ 
the Weyl group associated 
to the Lie algebra (\ref{eq:sscom}) consists of elements 
\begin{equation}
  \label{eq:Up}
  U_p:=e^{ip_\mu\hat x^\mu}\quad \text{with}\quad 
  \textrm{Tr}_{\ca{H}}\{U_p U^\dagger_{p'}\}=
     \frac{(2\pi)^{d/2}}{\textrm{Pf}(\theta)}\ \delta^d(p-p')\ ,
\end{equation}
where the trace is taken over the Hilbert space ${\cal{H}}$ and $\textrm{Pf}$ 
denotes the Pfaffian. The dimension $d=4$ in 
our case. The operator $U_p$ generates translations on the algebra 
$\ca{A}_\theta$, i.e. 
\begin{equation}
  \label{eq:ux}
  [U_p,\hat x^\mu]=\theta^{\mu\nu}p_\nu\ U_p\quad .
\end{equation}
Ordinary commutative functions are mapped to operators by the Weyl map:
\begin{equation}
  \label{eq:wmap}
  f(x)\longmapsto \hat f(\hat x)
      =\int\frac{d^dx\ d^dp}{(2\pi)^d}\ f(x)\ e^{-ipx}U_p\quad .
\end{equation}
Under the Weyl map (\ref{eq:wmap}) the star-product of commutative 
functions becomes the ordinary product of operators. 
Differentiation and integration become the following operations on operators 
acting on  ${\cal{H}}$:
\begin{equation}
  \label{eq:intdif}
  \partial_\mu\ \longmapsto\  -i(\theta^{-1})_{\mu\nu}[\hat x^\nu,\ .\ ]
  \quad ,\quad \int\! d^d x\ \longmapsto \ 
           (2\pi)^{d/2}\ \textrm{Pf}(\theta)\textrm{Tr}_{\ca{H}}\quad .
\end{equation}
\noindent
{\bf{Fock space.}} Let us now focus on $\mathbb{R}^4_\theta$. 
The matrix $\theta^{\mu\nu}$ in (\ref{eq:sscom}) is 
antisymmetric. Thus by an orthogonal transformation it can be brought to 
the canonical form
\begin{equation}
  \label{eq:theta}
  \left(\theta^{\mu\nu}\right)= 
          \begin{pmatrix}&-\theta_1&&\\\theta_1&&&\\&&&-\theta_2
                \\&&\theta_2&\end{pmatrix}\ ,
\end{equation}
where due to  of the above mentioned transformation to Cartesian 
coordinates w.r.t. 
the open string metric (\ref{eq:osm}) the noncommutativity parameters 
are related to
the $B$-field components used in section 5 of \cite{Seiberg:1999vs} as follows:
\begin{equation}
  \label{eq:thtab}
  \theta_i=2\pi\alpha' b_i\quad .
\end{equation}
In the following we assume that $\theta_{1,2}>0$. 
We will discuss the regime of
these parameters in more detail below. The form  
of $\theta^{\mu\nu}$ suggests
a natural decomposition of the noncommutative space 
into noncommutative planes
reflecting the residual isometries in the presence of a  
$B$-field background.
On each plane we introduce complex coordinates\footnote{In the following we omit
the indication of operators by hats. 
We follow the notation of \cite{Gross:2000ph}.}, 
 $z_1=x^1-ix^2\ ,\ z_2=x^3-ix^4$
such that the gauge fields write as
\begin{equation}
  \label{eq:Az}
  A_{z_1}=\hal(A_1+iA_2)\quad,\quad \bar A_{\bar z_1}=\hal(A_1-iA_2)
  =-(A_{z_1})^\dagger,
\end{equation}
and analogously for the second plane parametrised by $z_2$. After introducing the 
the operators 
\begin{equation}
  c_\alpha:=\fr{1}{\sqrt{2\theta_\alpha}}\ z_\alpha\quad,\quad
   c_\alpha^\dagger:=\fr{1}{\sqrt{2\theta_\alpha}}\ \bar z_\alpha\quad,\quad
\end{equation}
the algebra (\ref{eq:sscom}) becomes a two-oscillator algebra
and one can build up the Fock space as follows:
\begin{equation}
  \label{eq:fock}
  [c_\alpha,c^\dagger_\beta]=\delta_{\alpha\beta}\quad ,\quad 
  {\cal{H}}= {\cal{H}}_1\otimes{\cal{H}}_2\quad,\quad
  {\cal{H}}_\alpha=\overset{\infty}{\underset{n=0}{\bigoplus}}
  \ \mathbb{C}|n\rangle_\alpha\ ,
\end{equation}
where $|n\rangle_\alpha$ are the usual occupation number states.

With the identification (\ref{eq:intdif}) the covariant derivatives
in complex coordinates are the represented by the operators
\begin{equation}
  \label{eq:cov}
  \nabla_{z_\alpha}=
  -\frac{1}{\sqrt{\xi_\alpha}}\ c_\alpha^\dagger+A_{z_\alpha}\quad,\quad
  \bar\nabla_{\bar z_\alpha}=\frac{1}{\sqrt{\xi_\alpha}}
  \ c_\alpha+\bar A_{\bar z_\alpha}
  =-(\nabla_{z_\alpha})^\dagger\ ,
\end{equation}
where we have introduced the abbreviation $2\theta_\alpha=:\xi_\alpha$. Because
of noncommutativity and (\ref{eq:intdif}) the covariant 
derivatives for all  
fields in the adjoint representation act as 
$\nabla_{z_\alpha}\Phi:=[\nabla_{z_\alpha},\Phi]$, even 
if the gauge group is taken to be $U(1)$. The field strength in terms of
the derivatives (\ref{eq:cov}) writes as
\begin{eqnarray}
  \label{eq:fst}
  &&F_{z_\alpha,\bar z_\alpha}=[\nabla_{z_\alpha},\bar\nabla_{\bar z_\alpha}]
       \  -\frac{1}{\xi_\alpha} \nonumber\\
 &&F_{z_1,z_2}=[\nabla_{z_1},\nabla_{z_2}]\quad,\quad
   F_{z_1,\bar z_2}=[\nabla_{z_1},\bar\nabla_{\bar z_2}]\quad .
\end{eqnarray}

\section{A self-dual descent}

In the following we consider a gauge field 
\begin{equation}
  \label{eq:AM}
  A_M=(A_0,A_\mu,X_m)\quad , \quad \mu=(1,2,3,4)\ \ ,
\end{equation}
in ten or six dimensions, respectively. This  
depends on the supersymmetry content which one considers in four dimensions, 
obtained after dimensional reduction. 
As  is well known \cite{Brink:1977bc} the dimensional 
reduction of ${\cal{N}}=1$ super Yang-Mills theory from ten/six to 
four dimensions leads to 
${\cal{N}}=4$/${\cal{N}}=2$ super Yang-Mills theory. We describe two objects 
simultaneously. The five brane, which is an instanton solution embedded 
as a soliton 
in  the 4+1 dimensional theory contained in ten-dimensional super 
Yang-Mills theory. 
It  carries a 
five-form charge \cite{Abraham:1990nz} and corresponds to a $D0$-$D4$ 
brane system. 
The second 
object is an instanton solution of Euclidean four-dimensional 
Yang-Mills theory also embedded in 
(possibly Euclidean) ten-dimensional super Yang-Mills theory.
%\footnote{The dimensional 
%reduction of $10D$ SYM to Euclidean 4D SYM 
%can be done either according to $SO(10)\rightarrow S0(4)\times SO(6)$ 
%or according
%to $SO(9,1)\rightarrow S0(4)\times SO(5,1)$, where the second group 
%factor denotes 
%the resulting internal symmetry group. In fact, depending on the 
%convention for for 
%the dimensional reduction of the fermions there exists even more 
%different possibilities for 
%the resulting internal symmetry group. See e.g. \cite{Brink:1977bc} 
%for the reduction to 4D Minkowski space.}.     
The instanton solution of Euclidean four-dimensional Yang-Mills theory 
corresponds to a $D(-1)$-$D3$ brane system. In 
this case the quantum theory is well defined. This is possibly also true
on noncommutative space, since it seems that the 
$\theta$-deformation does not spoil the renormalisation of divergences 
due to planar loop graphs 
\cite{Grosse:2000yy,Bonora:2000ga,Sheikh-Jabbari:1999iw}. For supersymmetric 
Yang-Mills theories also the infrared divergences (UV/IR mixing) due to non-planar
loop contributions are under  much better control, see e.g. 
\cite{Santambrogio:2000rs,Zanon:2000nq}.  

To get down to the physically relevant $3+1$ dimensions in the 
case of the five brane one would also have to compactify one of the nontrivial 
directions. Thus the five brane is more correctly described by a caloron 
than an 
instanton. But we will not discuss this case here. Anyway, the two systems,  
$D0$-$D4$ and $D(-1)$-$D3$, are formally very similar and we will 
usually use the 
notation for the five brane but use the term instanton. To obtain
the Euclidean four-dimensional case one just has to set the temporal 
component
of the gauge field, classical and quantum, as well as time derivatives 
identically to zero in the following sections. 
For classical solutions this just corresponds to 
static solutions in the temporal gauge $A_0=0$.

We briefly review further dimensional reductions to show how lower 
dimensional
objects like monopoles/fluxons  and vortices are contained as 
special cases in 
the following relations. For the rest of the paper we indicate 
classical configurations
with caligraphical letters. Covariant derivatives w.r.t. 
classical background gauge fields
are denoted by $D_\mu$ instead of $\nabla_\mu$.\\

\noindent
{\bf{The instanton on $\mathbb{R}^4_\theta$.}}
In this case the non-vanishing components of the 
gauge field ${\cal{A}}_M$ are 
${\cal{A}}_\mu$ which depend only on the noncommutative 
coordinates $x^\mu$. For the
five brane this corresponds to static solutions in  temporal gauge. 
The noncommutative instanton  is then described by the 
four-dimensional self-duality equations\footnote{\label{fn}Nowadays  the term
instanton is usually used for anti-self-dual solutions. For linguistic  
simplicity and 
to avoid some signs we consider self-dual solutions. 
From the viewpoint of the geometry of $K3$ Calabi-Yau 
manifolds anti-self-dual fields appear to be more convenient.
But the  
results that we obtain are mutatis mutandis valid for anti-self-dual backgrounds. 
Correctly, self-dual solutions are 
identified with anti-$D0$-$D4$ branes but we keep 
our ``inexact'' notion.} $\ast {\cal{F}}={\cal{F}}$. In 
Cartesian coordinates they are 
\begin{equation}
  \label{eq:sd1}
  {\cal{F}}_{\mu\nu}=\hal\ \varepsilon_{\mu\nu\alpha\beta}
   {\cal{F}}^{\alpha\beta} .
\end{equation}

\noindent
{\bf{The monopole on $\mathbb{R}^2_\theta\times\mathbb{R}$.}}
Now we write the nontrivial gauge field as ${\cal{A}}_\mu
  =({\cal{A}}_{i=1\dots 3},\phi)$ 
and let the coordinates $x^{3,4}$ be commutative. Then the dimensional reduction
w.r.t. to  the $x^4$-direction in the self-duality equations (\ref{eq:sd1})
leads to:
\begin{equation}
  \label{eq:mp1}
  {\cal{F}}_{ij}=\ve_{ijk}D_k\phi . 
\end{equation}
These are the equations for the static monopole, again in temporal gauge.\\

\noindent
{\bf{The vortex on  $\mathbb{R}^2_\theta$.}} 
Reducing further the Bogomolnyi equations (\ref{eq:mp1}) with respect to the 
commutative coordinate $x^3$ one obtains the vortex equation 
on the noncommutative
plane. Writing the nontrivial gauge field components as 
${\cal{A}}_\mu=({\cal{A}}_{a=1,2},\phi_1,\phi_2)$ and introducing 
the complex field 
$\phi=\hal(\phi_1+i\phi_2)$ one obtains from (\ref{eq:mp1}):
\begin{equation}
  \label{eq:vo1}
  {\cal{F}}_{ab}=2i\ \ve_{ab}[\phi,\bar{\phi}]\quad ,\quad D_{-}\phi=0\quad ,
\end{equation}
where bar means complex conjugation and $D_{-}=\hal(D_1+iD_2)$. These
are (2+1)-dimensional static vortex equations in the temporal gauge. They differ from
 the noncommutative ANO-vortex equations which include a Fayet-Iliopoulos term
\cite{Popov:2003xg}.\\

In the following we will exclusively treat the instanton on 
$\mathbb{R}^4_\theta$ but a number  of the formal manipulations 
 will be valid  also for the 
lower-dimensional objects through the identification we gave above.

\subsection{The small instanton singularity regime}

In what follows  we restrict ourself to a single
$D4$-brane, i.e. the gauge group is $U(1)$. Thus 
the nontrivial solutions (instantons) that
we will consider have no counterpart in the commutative theory.
The physical properties of these solutions and the possible
quantum states will crucially depend on the properties of the 
noncommutative parameters $\theta^{\mu\nu}$. The most important
characteristic is the sign of the Pfaffian $\textrm{Pf}(\theta)=\theta_1\theta_2$. 
In the case that $\textrm{Pf}(\theta)>0$ $\theta^{\mu\nu}$ is 
continuously connected to  the self-dual point $\theta^{\mu\nu}_{SD}$, 
where $\theta_1=\theta_2$. Therefore we call this the self-dual regime
of $\theta^{\mu\nu}$. For  $\textrm{Pf}(\theta)<0$ the situation is 
reversed and $\theta^{\mu\nu}$ is continuously connected to the 
anti-self-dual point $\theta^{\mu\nu}_{ASD}$, where $\theta_1=-\theta_2$.
In the past several authors constructed $U(1)$ instanton solutions
on $\mathbb{R}^4_\theta$ with opposite duality for the field strength
${\cal{F}}_{\mu\nu}$ and the noncommutativity parameter $\theta^{\mu\nu}$
\cite{Chu:2001cx,Nekrasov:1998ss,Nekrasov:2000zz,Kraus:2001xt}. 
For the rest of the paper we will concentrate on the opposite case. We assume
 $\theta^{\mu\nu}$ to be in the self-dual regime, i.e. $\textrm{Pf}(\theta)>0$,
where we approach the special point $\theta^{\mu\nu}_{SD}$ from values
$\theta_1>\theta_2$. The field strength ${\cal{F}}_{\mu\nu}$ will 
be either self-dual
for general $\theta^{\mu\nu}$ in the self-dual regime or become self-dual
in the special point $\theta^{\mu\nu}_{SD}$. 
The latter case corresponds to the small instanton singularity.

\section{Unstable solitonic states}

Decomposing the gauge field according to (\ref{eq:AM}) the 
dimensionally reduced 
(super) Yang-Mills action (\ref{eq:ncYM1}) writes as 
{\setlength\arraycolsep{2pt}
\begin{eqnarray}
  \label{eq:ymrd}
  S=\frac{1}{4g_{YM}^2}\int 
  \textrm{Tr}&& \{(F_{\mu\nu})^2+2(\nabla_\mu X_m)^2
                 +[X_m,X_n]^2 \nonumber\\
     &&-2(\nabla_0 X_m)^2-2(\nabla_\mu A_0-\partial_t A_\mu)^2\}
       +  \textrm{fermions} \ \ ,
\end{eqnarray}}
where we have introduced the abbreviation 
\begin{equation}
  \label{eq:tr}
  \int\textrm{Tr}:=\int\! dt\ (2\pi)^2\textrm{Pf}
(\theta)\textrm{Tr}_{{\cal{H}}}\quad .
\end{equation}
When restricting to the (Euclidean) four-dimensional case one has to omit the 
integration over time in (\ref{eq:tr}).
For later convenience we write down  the fermionic part of the action using the  
higher dimensional notation:
\begin{equation}
  \label{eq:lferm}
  S_{\textrm{ferm}}=
  -\frac{1}{2g_{YM}^2}\int\textrm{Tr}\{\bar\lambda\Gamma^M\nabla_M\lambda\}\quad,
\end{equation}
where $\bar\lambda=\lambda^\dagger  \Gamma^0$. For most of our considerations
we will assume $\ca{N}=4$ supersymmetry\footnote{Here the specific name 
for the extended supersymmetries refers to the four-dimensional case.} so 
that $\lambda$ is a  
Majorana-Weyl spinor in ten-dimensional space-time. In the case of $\ca{N}=2$ 
supersymmetry
$\lambda$ is a Weyl spinor in six-dimensional space-time. The action is invariant under 
the same susy-transformations as in the commutative case\footnote{Note that 
the susy currents cannot be determined by local susy transformations. 
In this case the susy parameters became nontrivial operators  and
relevant manipulations  are no longer possible.}: 
\begin{eqnarray}
  \label{eq:susy}
  \delta A_M&=&\fr{1}{2}\left(\bar\lambda\Gamma_M\epsilon_1
    -\bar\epsilon_2\Gamma_M\lambda\right) \nonumber\\
  \delta\lambda &=&\fr{1}{2}F_{MN}\Gamma^{MN}\epsilon_1\quad,\quad 
  \delta\bar\lambda=-\fr{1}{2}\bar\epsilon_2 \Gamma^{MN}F_{MN}\quad,
\end{eqnarray}
where $\Gamma^{MN}=\fr{1}{2}[\Gamma^M,\Gamma^N]$ and the supersymmetry parameters 
are proportional to the unit operator. For $\ca{N}=4$ supersymmetry 
$\epsilon_1=\epsilon_2$. 
In the following we will mostly concentrate on the bosonic fields. The fermionic fields
are either determined by supersymmetry or put to zero if not stated differently. 
In the operator formalism the noncommutative 
$U(1)$ gauge symmetry of the action is represented by unitary 
transformations $U_{\ca{H}}$ acting on the  Hilbert 
space (\ref{eq:fock}):
\begin{equation}
  \label{eq:gtraf}
  A_\mu\rightarrow U^{-1}A_\mu U+U^{-1}\partial_\mu U\ \ ,\ \ 
  X_m\rightarrow  U^{-1}X_m U \quad,
\end{equation}
with $U^{-1}U=UU^{-1}=\unit$. The operator $\partial_\mu$ is given
in (\ref{eq:intdif}).

The equations of motion for static configurations in the temporal gauge 
are 
\begin{eqnarray}
  \label{eq:eom}
  D^\mu{\cal{F}}_{\mu\nu}+[D_\nu \ca{X}_m,\ca{X}_m]&=&0\quad, \nonumber\\
  \mD^2 \ca{X}_m+[[\ca{X}_m,\ca{X}_n],\ca{X}_n]&=&0\quad.
\end{eqnarray}
Here and in the following $\mD^2=D^\mu D_\mu$.
Self-dual gauge fields, for which 
$(\ast\ca{F})_{\mu\nu}:=\tilde{\ca{F}}_{\mu\nu}=\ca{F}_{\mu\nu}$, 
with non-vanishing scalar fields $\ca{X}_m$ are solutions to these
equations if the scalar fields $\ca{X}_m$ are \emph{commuting zero-modes}
of the operator $D_\mu$, i.e. 
\begin{equation}
  \label{eq:czm}
  D_\mu \ca{X}_m=0\quad ,\quad [[\ca{X}_m,\ca{X}_n],\ca{X}_n]=0\quad .
\end{equation}
If $[\ca{X}_m,\ca{X}_n]=0$ these commuting zero-modes do 
not contribute to the action (\ref{eq:ymrd}).
The (four-dimensional) classical action for such solutions is given by
\begin{equation}
  \label{eq:scl}
  S^{cl}=\frac{1}{4g^2_{YM}}\int\textrm{Tr}(\ca{F}_{\mu\nu}\ca{F}^{\mu\nu})\quad ,
\end{equation}
and therefore 
the commuting zero-modes (\ref{eq:czm}) are  moduli of self-dual solutions.
We will discuss this in more detail below.\\

\noindent
{\bf{Generating solutions.}} The so called solution generating technique 
generates nontrivial solutions from  vacuum configurations
\cite{Harvey:2000jb}. This method uses non-unitary isometries 
of the equations of motion to generate new configurations 
which are not gauge-equivalent to the vacuum. An important 
input in this context are the so called shift operators.
For the noncommutative $\mathbb{R}^4_\theta$ a shift operator
of order $k$ can be defined as a product of  
order one shift operators
{\setlength\arraycolsep{2pt}
\begin{eqnarray}
  \label{eq:shift}
   S^{\textrm{I}}&
        =&\sum_{m=0,n=1}\ket{m,n}\bra{m,n}+\sum_{m=0}\ket{m+1,0}\bra{m,0}\ \ ,
   \nonumber\\
  S^{\textrm{I}\!\textrm{I}}&
        =&\sum_{m=1,n=0}\ket{m,n}\bra{m,n}+\sum_{n=0}\ket{0,n+1}\bra{0,n}\ \ .
\end{eqnarray}}\noindent
These operator shift states $|m,n\rangle\in\ca{H}$
in the first and second occupation number, respectively.
Different sequences of $k$ factors of 
$S^{\textrm{I},\textrm{I}\!\textrm{I}}$ define different 
shift operators of order $k$ which are nevertheless gauge 
equivalent \cite{Furuuchi:2000vc}. The relevant properties of an order $k$ 
shift operator are 
\begin{equation}
  \label{eq:sk}
  S_k^\dagger S_k=\unit\quad ,
  \quad S_kS_k^\dagger=\unit-P_k\quad, \quad P_k S_k=0,
\end{equation}
where $P_k$ is a projection operator being the unit on the 
subspace $V_{k}=P_k\ca{H}\subset\ca{H}$ on which $S_k$ acts non-trivially,
so that\footnote{Drawing the states $|m,n\rangle$, $m,n=0,1,2\dots$, 
on a plane with lattice points $(m,n)$,  $\textrm{Tr}P_k$ is equal
to the ``area'' of the subspace $V_k$ in this picture.}  
\begin{equation}
  \label{eq:pk}
  \textrm{Tr} P_k=k.
\end{equation}
With this preparation one can immediately write down charge $k$ 
solutions generated
from the vacuum $D_\mu^{\textrm{vac}}= -i(\theta^{-1})_{\mu\nu} x^\nu$ 
and the associated field strength \cite{Aganagic:2000mh}\footnote{For similar 
constructions of $U(1)$ instantons in the signature $2+2$ see \cite{Ihl:2002kz}.}:
\begin{equation}
  \label{eq:dcart}
  D_\mu=iS_k(\theta^{-1})_{\mu\nu}x^\nu S_k^\dagger\quad ,\quad
  \ca{F}_{\mu\nu}=-i(\theta^{-1})_{\mu\nu}\ P_k\quad.
\end{equation}
For self-dual $\theta_{\mu\nu}$ the field strength (\ref{eq:dcart}) 
is self-dual. Expressed in complex coordinates this solution writes as
\begin{equation}
  \label{eq:dcomp}
  D_{\alpha}=\frac{1}{\sqrt{\xi_\alpha}}\ S_k c_\alpha^\dagger S_k^\dagger
  \quad ,\quad 
  \bar D_{\alpha}=-\frac{1}{\sqrt{\xi_\alpha}}\ S_k c_\alpha S_k^\dagger
  \ ,
\end{equation}
where $\alpha=1,2$ refers to $z_{1,2}$. The classical action 
(\ref{eq:scl}) or mass  and
the instanton charge for this solution are easily obtained as 
(here the time-integral is not included in $\int \textrm{Tr}$) 
{\setlength\arraycolsep{2pt}
\begin{eqnarray}
  \label{eq:top}
  M_{5}&=&
  \frac{4\pi^2\ k}{2 g^2_{YM}}\ \left(\frac{\theta_1}{\theta_2}+
   \frac{\theta_2}{\theta_1}\right)\ \ ,\nonumber\\
Q&=&-\frac{1}{16\pi^2}\int \textrm{Tr}F_{\mu\nu}(\ast F)^{\mu\nu}=k \ \ ,
\end{eqnarray}}
where $M_{5}$ denotes the classical mass of the
five brane. In fact the notion of mass is somewhat 
misleading in this case since, as we will see later, 
this configuration is not stable for generic 
$\theta_{\mu\nu}$. So it would be more appropriate to use the term energy.  
Note that at the self-dual point $\theta_1=\theta_2$ the 
classical five-brane mass and thus the classical noncommutative $U(1)$ instanton action
 equals half of the 
%\footnote{The reason for the 
%sign is that we work in 5+1 dimensions. The definition of the 
%Euclidean four dimensional action involves an additional sign compared 
%to the spatial part of the 5+1 dimensional action, see e.g.
%\cite{Belitsky:2000ws}.}
$k$-instanton action $S_{\textrm{inst}}=\frac{8\pi^2}{g^2_{YM}}k$ for (commutative) 
$U(2)$ instantons
 \cite{Belitsky:2000ws}.\\

\noindent
{\bf{D-brane interpretation.}} In \cite{Aganagic:2000mh} 
the D-brane interpretation of 
the solutions (\ref{eq:dcart}) was conjectured. This is  based on the matching of 
the (classical) binding energies for the five brane  and the (anti) 
$D0$-$D4$ brane system. In the noncommutative Yang-Mills description the binding energy 
for a single $D0$-brane is given by 
\begin{equation}
  \label{eq:Eb}
  E_{\textrm{bind}}=M_5-M_5^{BPS}=\frac{2\pi^2}{g_{YM}^2}
      \left( \sqrt{\frac{\theta_1}{\theta_2}}-\sqrt{\frac{\theta_2}{\theta_1}}\right)^2
          =\frac{1}{2 g_s\sqrt{\alpha'}}\left(\frac{1}{b_1}-\frac{1}{b_2} \right)^2 \quad,
\end{equation}
where in the second equation we have written the binding energy in 
terms of the string quantities (\ref{eq:thtab}) and (\ref{eq:ymst}) for $p=4$. 
In units where $\alpha'=1$ the mass defect of the formation of a $D0$-$D4$ bound 
state in the limit 
$b_i\gg 1$ is given 
by \cite{David:2000um}\footnote{Our notation is related to the one of \cite{David:2000um} 
as follows:
 $b_1=b$, $b_2=-b'$ and $g_s=2\pi^2g$. The range $b_i\gg 1$ of the $B$-field corresponds 
to ``case $(3)$'' in \cite{David:2000um}.}
\begin{equation}
  \label{eq:dm}
  \Delta M=M_{D0}+M_ {D4}-M_{D0+D4}
    =\frac{1}{2 g_s}\left(\frac{1}{b_1}-\frac{1}{b_2}\right)^2
  \quad,
\end{equation}
and thus matches exactly the binding energy (\ref{eq:Eb}). We will see below that 
also the excitation spectra of the solitonic solution (\ref{eq:dcart}) match the one of the $D0$-$D4$ 
system.

\subsection{Zero-mode moduli}

From the properties (\ref{eq:sk}) one 
immediately sees that the classical equations of motion 
(\ref{eq:eom}) are fulfilled by (\ref{eq:dcart}) and the projector $P_k$ projects on the 
space of zero-modes given by 
\begin{equation}
  \label{eq:zm}
  D_\mu\ca{X}=0 \textrm{ for } \ca{X}\in V_k\otimes V_k^\ast \quad .
\end{equation} 
Up to gauge-equivalence the space $V_k$ is spanned by the states
$|j\rangle:=|m,n\rangle$ with
\begin{equation}
  \label{eq:zmspace}
     j=m+\frac{(m+n)(m+n+1)}{2}\ \ ,\ \ j=0,\dots,k-1\ .
\end{equation}
So $V_k$ is the lower left corner of the $(m,n \geq0)$-plane with 
the ``area'' $\textrm{Tr}(P_k)=k$. Thus a general zero-mode 
is of the form
\begin{equation}
  \label{eq:gzm}
  \ca{X}=\sum_{j,j'=0}^{k-1} \ca{X}_{jj'}
  |j\rangle\langle j'|\quad .
\end{equation}
The solution (\ref{eq:dcart}) breaks the $U(\ca{H})$ gauge 
symmetry down to $U(k)$ which acts non-trivially  only in the 
subspace $V_k\otimes V_k^*$:
\begin{equation}
  \label{eq:uk}
  U(k)\ni U=\sum_{j,j'=0}^{k-1} U_{jj'}
    |j\rangle\langle j'|\quad .
\end{equation}
Stated differently, these are the transformations which leave
the gauge connections (\ref{eq:dcart}) invariant and are therefore 
the analogue of the stability group with 
corresponding gauge orientation zero-modes for commutative $U(2)$ 
instantons \cite{Belitsky:2000ws}.

The zero-modes (\ref{eq:zm}) transform under the residual 
gauge symmetry in the adjoint representation:
\begin{equation}
  \label{eq:gs}
  \ca{X}\rightarrow U^\dagger\ca{X}U=
  \sum_{i,j=0}^{k-1}(U_{i'i})^\dagger\ca{X}_{i'j'}U_{j'j}|i\rangle\langle j|
\end{equation}
where summation over repeated matrix indices is implied.
Of special interest are \emph{commuting}  zero-modes (\ref{eq:czm})
which are possible nontrivial scalar field solutions. The simplest
solution to (\ref{eq:czm}) is given by diagonal matrices in the 
representation (\ref{eq:gzm}):
\begin{equation}
  \label{eq:scalars}
  (\ca{X}^m)_{jj'}
  = \textrm{diag}(\alpha^{(m)}_0,\dots,\alpha^{(m)}_{k-1}).
%  \begin{pmatrix}0\\&\ddots\\&&\alpha_m\\&&&\ddots\\&&&&0\end{pmatrix}
\end{equation}
We will see later that the vectors $\alpha_j^{m}$ with components 
in the $m$ directions transverse to the $D4$-brane can be interpreted
as the positions of the $k$ $D0$-branes outside the $D4$-brane.
We also see that if the $k$ $D0$-branes are separated, i.e. 
$\alpha^{(m)}_0\neq\dots\neq\alpha^{(m)}_{k-1}$, the residual $U(k)$ gauge symmetry
is further broken down to $U(1)^k$.

In general the (second) equation in (\ref{eq:czm}) allows more nontrivial solutions 
for $k\times k$ matrices. In fact these are the equations for static bosonic 
configurations of the BFSS matrix quantum mechanics which has been conjectured to 
be  equivalent to M-theory \cite{Banks:1996vh}. Nontrivial 
solutions to (\ref{eq:czm}) may for example describe $D0$-$D4$ bound states
without a $B$-field background \cite{Massar:1999jp}. 
We will give a more detailed description of the relation to the 
matrix model below.

\subsection{Fluctuation spectrum}

In this section we investigate the fluctuation spectrum or equivalently the 
asymptotic
states of the quantum field theory in the presence of
the above discussed background solutions. For this we expand the full fields
around the classical background (\ref{eq:dcart}):
\begin{equation}
  \label{eq:exp}
  A_\mu\rightarrow\ca{A}_\mu+a_\mu\quad , \quad A_0\rightarrow a_0\quad ,\quad 
  X_m\rightarrow\ca{X}_m+X_m\ \ .
\end{equation}
In addition we need a gauge fixing condition. As  has been shown in the 
quantisation of commutative monopoles and vortices the most comfortable gauge in the 
presence of classical backgrounds  is the covariant background gauge 
implemented in the higher dimensional space \cite{Rebhan:2003bu,Rebhan:2004vn}:
\setlength\arraycolsep{2pt}
\begin{eqnarray}
  \label{eq:gf}
  S_{\textrm{gf}}&=&\frac{1}{2g_{YM}^2}\int\textrm{Tr}(D^M a_M)^2 \nonumber\\
   &=&\frac{1}{2g_{YM}^2}
   \int\textrm{Tr}(-\dot{a}_0+D_\mu a_\mu+[\ca{X}_m,X_m])^2\quad.
\end{eqnarray}
By the usual procedure which can be applied without any problems to the 
noncommutative
case one obtains the ghost action associated to (\ref{eq:gf}):  
\begin{equation}
  \label{eq:ghost}
  S_{\textrm{FP}}=
 -\frac{1}{g_{YM}^2}\int\textrm{Tr}(b D^M D_M c+ b D^M[a_M,c])\ \ .
\end{equation}
%As in the commutative case ghosta nd gauge fixing action can be written as
%BRST-transformation of a gauge-fixing fermion \cite{}:
%\begin{equation}
%  \label{eq:brst1}
%  \ca{L}_{\textrm{gf+FP}}=\delta_{BRST}\Psi\quad\textrm{with}\quad 
%  \Psi=\frac{2}{g_{YM}^2}}
%\end{equation}
We give here the ghost action for future reference, since it will be needed 
for one-loop corrections.

We  focus on the $k=1$ background, such that 
\begin{equation}
  \label{eq:P}
  P_{k=1}=:P=|00\rangle\langle00|\quad .
\end{equation}
Without loss of generality we
can choose the $D0$-brane to be separated in the $m=1$ direction
from the $D4$-brane. Thus the only nontrivial scalar field is 
\begin{equation}
  \label{eq:fi1}
  \ca{X}_1=i\mu P\quad ,
\end{equation}
where $\mu\in \mathbb{R}$ since all fields are anti-hermitian. 
Including the gauge-fixing action (\ref{eq:gf}) the  
action quadratic in the fluctuations reads as
{\setlength\arraycolsep{2pt}
\begin{eqnarray}
  \label{eq:s2}
  S^{(2)}&=&\frac{1}{2g^2_{YM}}\int\textrm{Tr}\ 
    a_\mu\left\{\delta^{\mu\nu}(\partial_t^2-\mD^2+\mu^2[P,[P,.\ ]])
      -2[\ca{F}^{\mu\nu},.\ ]\right\}a_\nu\\
  &+&\frac{1}{2g^2_{YM}}\int\textrm{Tr}X_m\left\{(\partial_t^2-\mD^2)
    +\mu^2[P[P,.\ ]]\right\}X_m\nonumber\\
  &-&\frac{1}{2g^2_{YM}}\int\textrm{Tr}\ a_0\left\{(\partial_t^2-\mD^2)
    +\mu^2[P[P,.\ ]]\right\}a_0+\textrm{ghosts}+\textrm{fermions.}
  \nonumber
\end{eqnarray}}
%A technical remark is in order: To obtain the above form of the quadratic action
%one has to use the cyclicity under the trace several times. In an infinite-dimensional
%space this is not true for  general operators. But per definition 
%the full quantum fields, not necesary the individual modes, 
%are Hilbert-Schmidt operators.
%Whereas classical backgrounds are in general bounded op     
In the following we will mostly concentrate on the bosonic fluctuations.
The fermionic ones can be determined by supersymmetry, at least at
the self-dual point $\theta_1=\theta_2$. 
The fluctuation operators for the different fields are very similar,
except for the additional spin-coupling
through the background field $\ca{F}^{\mu\nu}$ for the ``gluons'' $a_\mu$.
The scalar field background (\ref{eq:fi1}), i.e. the separation of the
$D0$-brane,  induces a mass term for all 
quantum fields.
The $a_0$ fluctuations are not physical and will be cancelled by the
ghost.  But let us first look at the scalar fields.\\

\noindent
{\bf{Scalar fluctuations.}} We look for the spectrum (discrete and 
continuous) of the fields $X_m$ and $a_0$. Thus we have to solve  the 
eigenvalue problem
\begin{equation}
  \label{eq:sop}
  -\Delta\vf_\omega=\omega^2\vf_\omega\quad \textrm{with}\quad
  -\Delta=-\mD^2+\mu^2[P,[P,\ .\ ]]\ \ .
\end{equation}
The string picture as well as the form of the background 
(\ref{eq:dcart}), (\ref{eq:dcomp}) suggest the separation of the
eigenmodes $\vf_\omega$ into different sectors of the Hilbert space
$V=\ca{H}\otimes\ca{H}^*$:
\begin{equation}
  \label{eq:fl1}
  \vf_\omega:=\alpha P+P a S^\dagger+ S b P+SdS^\dagger\quad ,
\end{equation}
where $\alpha\in \mathbb{C}$ and $a,b,d$ are general operators $\in V$.
In the following we choose the shift operator 
$S=S^{\textrm{II}}$ in (\ref{eq:shift}).
Equation (\ref{eq:fl1}) corresponds to a decomposition 
$V=V_{00}\oplus V_{04}\oplus V_{40}\oplus V_{44}$ according to
open strings on the $D0$-brane, stretched between the $D0$-$D4$-brane and
open strings living on the $D4$-brane. This is quite analogous to the 
noncommutative $\mathbb{R}_\theta^2$ as it was  considered in 
\cite{Aganagic:2000mh,Gross:2000ph}. 
But as we will see, the slightly 
different definition of the sectors in (\ref{eq:fl1}) will simplify
the analysis especially for more involved situations which we will 
consider below as well as for higher dimensional noncommutative 
spaces.  Compare also the appendix of \cite{Rangamani:2001cn}.
Recalling the property (\ref{eq:sk}) one can see that $V_{00}$ and
$V_{44}$ do not feel the mass term in (\ref{eq:sop}). 
With the decomposition (\ref{eq:fl1}) one immediately 
obtains the spectrum of the $V_{00}$ and
the $V_{04,40}$ sectors. The eigenvalue equation (\ref{eq:sop}) 
for the $V_{44}$ sector in the background (\ref{eq:dcomp}) becomes
\begin{equation}
  \label{v44}
  2\sum _{\alpha=1,2}\frac{1}{\xi_\alpha}\left 
([c_\alpha,[c_\alpha^\dagger,d\ ]]+[c_\alpha^\dagger,[c_\alpha,d\ ]]\right)
   =\omega^2 d \quad .
\end{equation}
As expected this is just the free Laplacian since the 
 background (\ref{eq:dcomp}) is related by an isometry to the 
vacuum. Using the fact that $U_p$  generates 
translations (\ref{eq:ux}) and introducing the complex momenta 
$k_{\alpha=1,2}:=\halb(k_{1,3}+i\ k_{2,4})$ one can easily see that
the anisotropic noncommutative plane wave (\ref{eq:Up})
\begin{equation}
  \label{eq:ncpw}
 d=U_k=e^{i\sum_{\alpha}\sqrt{\xi_\alpha}
    (k_\alpha c_\alpha+\bar k_\alpha c^\dagger_\alpha)}\quad,
\end{equation}
is a solution to (\ref{v44}) with eigenvalue $\omega^2=k_\mu k^\mu$.
Collecting things together we have the following spectrum of
scalar fluctuations ($m,n\geq0$)
{\setlength\arraycolsep{2pt}
\begin{eqnarray}
  \label{eq:scsp}
  V_{00}:\quad\omega^2&=&0 \hspace{3,2cm},\quad \vf_0\ \ \  =P\ \ ,\nonumber\\
  V_{04}:\quad\omega^2&=& \fr{2m+1}{\theta_1}+\fr{2n+1}{\theta_2}+\mu^2
  \quad,\quad\vf_{m,n}=|0,0\rangle\langle mn|S^\dagger\ \ ,\nonumber\\
  V_{44}:\quad\omega^2&=&k_\mu k^\mu \hspace{2,6cm} ,\quad \vf_k
       \ \ \   =\fr{\sqrt{\theta_1\theta_2}}{2\pi}\ S U_kS^\dagger 
       \ \ . 
\end{eqnarray}}\noindent
The $V_{40}$ sector is simply given by the hermitian conjugate of the
$V_{04}$ states. The fluctuations are ortho-normalised w.r.t. the norm
$\textrm{Tr}(\vf^\dagger_\omega\ \vf_{\omega'})$.\\

\noindent
{\bf{Gauge field fluctuations.}} The gauge fields $a_\mu$  are now 
governed by the following equation:
\begin{equation}
  \label{eq:gffl}
  -\Delta_{\mu\nu}\ a_\nu=\omega^2\ a_\mu \quad\textrm{with}\quad 
    -\Delta_{\mu\nu}=-\delta_{\mu\nu}\Delta+2i\theta^{-1}_{\mu\nu}[P,.\ ]
    \ \ ,
\end{equation}
where $\Delta$ was already defined in (\ref{eq:sop}) and we have 
used (\ref{eq:dcart}). The additional spin coupling acts only in the 
$V_{04,40}$ sector. Thus, up to polarisation vectors 
$\ve_\mu$,  the gauge fields $a_\mu$ 
of the $V_{00}$ and $V_{44}$ sector  are 
given by the states and eigenvalues (\ref{eq:scsp}). 
With the canonical form (\ref{eq:theta}) of $\theta^{\mu\nu}$ the operator 
$\Delta_{\mu\nu}$ becomes block-diagonal with the
two by two blocks
\begin{equation}
  \label{eq:bl}
  -\tilde\Delta_{\alpha=1,2}=\begin{pmatrix}-\Delta&
              \frac{i}{\xi_\alpha}[P,.\ ]\\-\frac{i}{\xi_\alpha}[P,.\ ]
              &-\Delta\end{pmatrix}\quad .
\end{equation}
Therefore the gauge fields $a_\mu$ can be decomposed into ``up/down''
polarisations. With the ansatz $a_\mu=\ve_\mu\vf_{m,n}$ one
easily obtains the $V_{04}$ spectrum for the gauge fields
{\setlength\arraycolsep{2pt}
\begin{eqnarray}
  \label{eq:04gf}
  \omega^{2\pm}_\uparrow= 
     \fr{2m+1}{\theta_1}+\fr{2n+1}{\theta_2}\pm\fr{2}{\theta_1}+\mu^2
     \quad &,&\quad
  a_{\mu\uparrow}^{\pm}=\vf_{m,n}\left ({{\vec\ve^{\pm}}\atop{0}}\right)
  \quad ,\nonumber\\
  \omega^{2\pm}_\downarrow= 
   \fr{2m+1}{\theta_1}+\fr{2n+1}{\theta_2}\pm\fr{2}{\theta_2}+\mu^2
   \quad &,&\quad
   a_{\mu\downarrow}^{\pm}=\vf_{m,n}\left ({{0}\atop{\vec\ve^{\pm}}}\right)
     \quad ,
\end{eqnarray}}
where  $\vf_{m,n}$ is given in (\ref{eq:scsp}) and the polarisation
vectors are $\vec\ve^\pm=(i,\pm 1)$. Thus, depending on the 
polarisation the mode energies are shifted compared to the 
scalar field fluctuations (\ref{eq:scsp}). The eigenvalues 
of the $V_{04}$ sector for the case that the scalar field background
is trivial, which means $\mu=0$ here, 
have been obtained in \cite{Rangamani:2001cn}. For a non-vanishing 
scalar field background, i.e. $\mu^2>0$, the mass of the 
$V_{04}$ fluctuations increases with $\mu^2$. On the other hand 
$\mu$ is the amplitude of the scalar field (\ref{eq:fi1}). 
This fits perfectly with the string interpretation of $\ca{X}_1$ 
as the separation of the $D0$- from the $D4$-brane. As the 
distance increases the energy for strings stretched between them also 
increases. 
%%%%%%%%%%%%%%%%%%%%%%%%%%%%%%%%%%%%%%%%%%%%%%%
%The full (quantum) fields $a_\mu$, $X_m$ have the expansion
%{\setlength\arraycolsep{2pt}
%\begin{eqnarray}
%  \label{eq:qf}
%  a_\mu
%  =i\alpha_\mu P+\fr{\sqrt{\textrm{Pf}(\theta)}}{2\pi}
%    \int\!\!\frac{d^4k}{(2\pi)^4}\ \alpha_\mu(k)\ e^{-i\omega t} 
%    \vf_k
% \nonumber \\ &+&
%  \sum_{\overset{m,n}{\scriptstyle{\lambda=\uparrow\downarrow,s=\pm}}}\!\!
%    \alpha_{\lambda s}(m,n)\ \ve_\mu^{\lambda s}
%  + \sum_{m,n}
%      \alpha_\mu({\scriptstyle{m,n}})\ e^{-i\omega t}
%    \vf_{m,n} \nonumber \\ 
%  X_m=
%  i\beta_m P+\fr{\sqrt{\textrm{Pf}(\theta)}}{2\pi}
%    \int\!\!\frac{d^4k}{(2\pi)^4}\ \beta_m(k)\ e^{-i\omega t} 
%    \vf_k
%    + \sum_{p,q}
%      \beta_m({\scriptstyle{p,q}})\ e^{-i\omega t}
%    \vf_{p,q}.
%\end{eqnarray}}
%In both cases one has to subtract the hermitian conjugated expansion 
%to obtain anti-hermitian fields. The mode energies $\omega$ are given
%by (\ref{eq:scsp},\ref{eq:04gf}). Before discussing the spectrum 
%let us see which states are physical. For this 

If we set $\mu=0$ and use the identification (\ref{eq:thtab})  
 the low-lying mode energies of the $V_{04}$ sector, 
i.e. $m,n=0$ in (\ref{eq:scsp}) and (\ref{eq:04gf}), coincide with 
the energies obtained from quantising open strings stretched between
the $D0$- and $D4$-brane \cite{Seiberg:1999vs}. 
%\footnote{The additional 
%overall factor $1/\alpha'$ that we obtain in this way is due to the 
%convention of the SW-limit for which $\alpha' b_i\rightarrow\tilde b_i$ 
%with $\tilde b_i$ being finite. We have not absorbed a factor $\alpha'$ 
%in our definition of the $B$-field.}. Especially one also obtains a tachyon in the 
%$V_{04}$ sector of the gauge field fluctuation. 
As mentioned above 
we consider the regime $\theta_1>\theta_2$ such that the tachyon 
sits in the ``down'' fluctuation:
\begin{equation}
  \label{eq:tach1}
  \omega^{2-}_\downarrow=\frac{1}{\theta_1}-\frac{1}{\theta_2}+\mu^2
  \ <\   0 \quad\ \  \textrm{if}\ \ \quad 
  \frac{\mu^2}{2}\ <\frac{\xi_1-\xi_2}{\xi_1\xi_2}\ \ .
\end{equation}
For later convenience we have again introduced the quantities
$\xi_\alpha:=2\theta_\alpha$. As we will see later on 
$\sqrt{\xi_1-\xi_2}$ is the size of the instanton for non-self-dual 
 $\theta_{\mu\nu}$. Thus if the $D0$-brane approaches the $D4$-brane 
below the ``relative'' instanton size $\frac{\xi_1-\xi_2}{\xi_1\xi_2}$
the system becomes unstable. Before commenting on the rest of the 
spectrum let us impose the gauge condition  
associated to the gauge-fixing action (\ref{eq:gf}). In doing so
one can see that the gauge field and scalar field zero-modes 
\begin{equation}
  \label{eq:zm1}
  \sim\ve_\mu P\  ,\ \sim\alpha_m P\ \in V_{00}
\end{equation}
are unaffected by the gauge condition. The zero-modes in the gauge field
fluctuations can be understood as the usual translational zero-modes 
associated to the position of the $D0$-brane or its projection
in the $D4$-brane. But note that translations in noncommutative directions
play a subtle role since they are equivalent to gauge transformations 
\cite{Gross:2000ph}.
The zero-modes in the scalar fields are 
additional possible zero-mode moduli as were discussed in 
(\ref{eq:scalars}). Thus they refer to translations of the $D0$-brane in 
directions transversal to the $D4$-brane.  They have no counterpart 
in the commutative $U(2)$ instanton, where the operator $-\mD^2$
is strictly positive (note that for self-dual $\theta_{\mu\nu}$ the background 
is self-dual). This is in one-to-one correspondence 
with the irreducibility of the field strength, i.e. that $F\chi=0$
has no non-trivial solution \cite{Belitsky:2000ws}. 
Obviously the  solution 
(\ref{eq:dcart}) does not satisfy this condition. 
This may also have some impact on index theorems
on noncommutative space. In \cite{Kim:2002qm} for example 
it was assumed that the operator 
$\mD^2$ has no zero-modes.

The gauge condition for the  continuous $V_{44}$ spectrum restricts the polarisations
to be transversal to the momenta, i.e. $-\omega \ve_0+\ve_\mu k_\mu =0$.
For the $V_{04,40}$ spectrum the gauge condition implies various relations 
between the expansion coefficients for these modes.

\subsection{Self-dual point and $T$-duality\label{sse:td}}

Let us have a closer look at the special point $\theta_1=\theta_2:=\theta$ in 
the parameter 
space where the background gauge field (\ref{eq:dcart}) becomes self-dual and 
thus describes an 
instanton. The size of the instanton $\sqrt{\xi_1-\xi_2}$ shrinks to zero in
 this limit. 
At the self-dual point of $\theta_{\mu\nu}$ the solution 
(\ref{eq:dcart}) thus describes a 
small instanton. Although the moduli space still has a small instanton
singularity at this point, due to noncommutativity the solution (\ref{eq:dcart}) 
is regular for self-dual $\theta_{\mu\nu}$. 
%This was also observed for the 
% noncommutative $U(2)$ instanton \cite{Furuuchi:2000dx}.
The spectrum in the self-dual point can also be obtained continuously by the
limit $\theta_{1,2}\rightarrow\theta$ from the spectra (\ref{eq:scsp}), (\ref{eq:04gf}).
The states do not change  except for those in the $V_{44}$ sector, which become isotropic 
noncommutative 
plane waves. The mode energies of the $V_{04,40}$ sector are changed as follows:
{\setlength\arraycolsep{2pt}
\begin{eqnarray}
   \label{sden}
     \omega^2&\rightarrow &\frac{2}{\theta}\ (m+n+1)+\mu^2\ \ ,\nonumber\\
     \omega^{\pm 2}_\uparrow&\rightarrow&\frac{2}{\theta}\ 
     (m+n+1\pm 1)+\mu^2\ \ ,\nonumber\\
     \omega^{\pm 2}_\downarrow&\rightarrow&\frac{2}{\theta}\ (m+n+1\pm 1)+\mu^2\ \ .
\end{eqnarray}}
The most important change in the spectrum is that now for arbitrary small 
separation $\sim\mu$ of the $D$-branes 
the spectrum is positive. The tachyon disappears. This was expected since
the gauge field background is self-dual. This is consistent with the fact  
that the $D0$-$D4$ system is 
supersymmetric and BPS for (anti)self-dual $B$-field backgrounds 
\cite{Seiberg:1999vs}. 

The up/down gauge field modes are degenerate because of the enhanced symmetry.
If we set $\mu=0$, i.e. when the $D0$-brane sits on the top of the $D4$-brane, 
in addition to the translational zero-modes (\ref{eq:zm1}) two independent
zero-modes in the $V_{04,40}$ sector appear. It is easy to see that they are not affected 
by the gauge condition.  We do not have an explicit interpretation in terms of moduli 
for these additional zero-modes. We can only suppose that they are related to 
the ``superconformal zero-modes''. In \cite{Kim:2002qm} it was shown that for self-dual 
noncommutative instantons with self-dual background $\theta_{\mu\nu}$
these zero-modes are unaffected by the noncommutativity. The reason for this is 
that the ADHM constraints are the same as for the  commutative case.

In the following we will assume that $\mu=0$.\\

\noindent
{\bf{The fluxon and T-duality.}} In \cite{Gross:2000ph} the so called fluxon  
solution to equation (\ref{eq:mp1}) and 
its spectrum were considered. The corresponding $D$-brane picture
is a tilted $D1$-brane piercing a $D3$-brane at an angle 
$\tan\psi=\frac{2\pi\alpha'}{\theta}$. The one-oscillator Fock space conventions
are obtained by setting $\theta_1=\theta$ and $\theta_2=0$. The 
single fluxon is given by ($D_z=:D,\ \bar D_{\bar z}=:\bar D$) 
\begin{equation}
  \label{eq:flux1}
  D=\fr{1}{\sqrt{\xi}}Sc^\dagger S^\dagger\quad,\quad
  \bar D=-\fr{1}{\sqrt{\xi}}Sc S^\dagger\quad,\quad
  \ca{A}_3=0\quad,\quad
  \phi=-\fr{2i}{\xi}\ x^3 P\quad,
\end{equation}
where the shift operator $S$ and the projector are now given by 
\begin{equation}
  \label{eq:flux2}
  S=\sum_{n=0}^{\infty}|n+1\rangle\langle n|\quad,\quad 
  P=|0\rangle\langle 0| \quad.
\end{equation}

Introducing $\ca{A}_\mu=(\ca{A}_{i=1,2,3}\ ,\phi)$ as suggested in (\ref{eq:mp1}) 
one finds the associated field strength $\ca{F}_{\mu\nu}$ to be of the same form
as the field strength  
(\ref{eq:dcart}) but with  self-dual $\theta_{\mu\nu}$ and the projector as 
given in (\ref{eq:flux2}).

The fluctuation spectrum of the fluxon, analogously to the $D0$-$D4$ system, 
decomposes into different sectors $V_{11}$, $V_{33}$, $V_{13,31}$ 
characterising strings living on the $D1$- or
the $D3$-brane and strings stretched between them. Analogously to (\ref{eq:scsp}) 
the scalar field eigenmodes $X_m$, except for fluctuations of the
nontrivial Higgs field $\phi$ in (\ref{eq:flux1}),  are eigenmodes 
of $-\mD^2$ ($m,n\geq0$):
{\setlength\arraycolsep{1pt}
\begin{eqnarray}
  \label{eq:flux3}
  V_{11}:\quad\omega^2=k^2_3 \hspace{2.3cm}&,&\quad \vf_0=e^{ik_3x^3}P \nonumber\\
  V_{13}:\quad\omega^2=\fr{2}{\theta}(m+n+1)\quad&,&
   \quad\vf_{mn}=e^{-x_3^2/\xi}H_m(\fr{x_3}{\sqrt{\theta}})|0\rangle\langle n|S^\dagger
                                                            \nonumber\\
  V_{33}:\quad\omega^2=\vec k^2\hspace{2.2cm}&,&\quad\vf_{k}=
              e^{ik_3x^3}Se^{i\sqrt{\xi}(kc+\bar kc^\dagger)}S^\dagger\ \ .
\end{eqnarray}}
Here $H_m$ are the Hermite polynomials and we have omitted 
proper normalisation constants here (see \cite{Gross:2000ph} 
for details).
The $V_{11}$ fluctuation becomes a normalisable zero-mode (per unit length) 
for vanishing momentum $k_3$, describing translation moduli transverse to the
$D1$-string \cite{Gross:2000ph}\footnote{Strictly speaking  
these zero-modes are not normalisable 
because of the $x_3$-integration. But having in mind that the 
$D1$-string will be $T$-dualised the motion along the $D1$-string is 
constrained to a point and so is its projection on the $x_3$ axis. Thus the plane wave 
in the $x_3$ direction might become a normalisable state as  the 
one in the $V_{44}$-sector. See the following discussion.}. 
They are analogous to zero-modes that 
  we found  in the $V_{00}$ sector of  the $D0$-$D4$ system 
(\ref{eq:scsp}). The $V_{13,31}$ spectrum is the same as the $V_{04,40}$ 
spectrum at the self-dual point, i.e. the first line in  (\ref{sden}). 
As for the  $D0$-$D4$ system there is a massless continuum  in the 
$D1$-$D3$ system but the degeneracy is different.

Grouping also the  gauge field and Higgs field fluctuations $\eta$
as $a_\mu=(a_{i=1,2,3}, \eta)$ one obtains the fluctuations analogously to
(\ref{eq:04gf}). As mentioned above, the background field strength 
$\ca{F}_{\mu\nu}$ is of the same form as for the instanton at the self-dual
point and thus the fluctuation operator for the modes $a_\mu$ is of the form
(\ref{eq:gffl}) with $\theta_1=\theta_2=\theta$. Hence the eigenmodes 
are decomposed according to the same polarisations as in (\ref{eq:04gf}). 
The mode energies are given by (\ref{sden}).

The equivalence of the spectra of the $D1$-$D3$ system  and the $D0$-$D4$ 
system  at the self-dual point could  probably be understood from $T$-duality.
By $T$-dualising the direction along the $D1$-brane the $D1$-brane becomes a 
$D0$-brane whereas the in this respect tilted $D3$-brane turns into a 
$D4$-brane with and additional magnetic field on it. One thus ends up with a 
$D0$-$D4$ system with a rank four $B$-field. Hence, in the presence of a 
$B$-field $T$-duality transformations also act on  the $B$-field 
background \cite{Seiberg:1999vs}.  This way a noncommutative 
theory can be mapped to
a commutative or partial commutative theory as  we observe it here,
where $\mathbb{R}_\theta^4\rightarrow\mathbb{R}_\theta^2\times\mathbb{R}$.
At the classical level we observe that the number of zero-mode moduli
is the same in both cases\footnote{There are five  scalars $X_m$ in both
cases.}. On the other hand the so called Nahm-duality identifies the
moduli spaces of self-dual solutions on dual four-tori 
\cite{Hamanaka:2003cm,Schenk:1986xe,Braam:1988qk}. 
Different limits for the radii of the tori lead to different non-compact
spaces as for example $\mathbb{R}^4$. This duality is also reflected 
in the ADHM construction of instantons (see below). 

Here one finds similar identifications when $T$-dualising single directions
and not only for the moduli space of self-dual configurations 
but also for the fluctuation spectrum. We will comment on this later when considering the
ADHM construction, but we have to leave a more detailed investigation of this issue 
 for future studies.  
\\

\noindent
{\bf{Supersymmetry of the spectrum of BPS backgrounds.}}
We briefly discuss a property of the fluctuation spectrum in the presence
of a self-dual background which will become important below.  
Introducing the generators of the quarternionic algebra,   
$\sigma^\mu=(\vec\sigma,-i\unit)$ and $\bar\sigma^\mu=(\vec\sigma,i\unit)$,
one can define the self-dual and anti-self-dual projector
\begin{eqnarray}
  \label{sdasd}
  \sigma^{\mu\nu}:=\fr{1}{4}(\sigma^\mu\bar\sigma^\nu-
                              \sigma^\nu\bar\sigma^\mu)\quad,\quad
  \bar\sigma^{\mu\nu}:=\fr{1}{4}(\bar\sigma^\mu\sigma^\nu-
                                  \bar\sigma^\nu\sigma^\mu)\quad . 
\end{eqnarray}
Writing the background covariant derivatives in spinor notation,
$\slashed{D}:=\sigma^\mu D_\mu\quad$ and 
$\slashed{\bar D}:=\bar\sigma^\mu D_\mu$ , 
one has for a self-dual background the following two quadratic operators:
\begin{equation}
  \label{eq:quop}
  \slashed{D}\slashed{\bar D}=\mD^2+\sigma^{\mu\nu}[\ca{F}_{\mu\nu},.\ ]\quad,
  \quad \slashed{\bar D}\slashed{D}=\mD^2\quad,
\end{equation}
where we have used that $\bar\sigma^{\mu\nu}\ca{F}_{\mu\nu}=0$ for a self-dual
background. The factorisation property (\ref{eq:quop}) 
implies that also on the noncommutative space the two operators in (\ref{eq:quop}) 
are isospectral except for zero-modes. 
But as in the commutative case the spectral 
densities of the continuous spectra are in general different\footnote{
In general we include the continuous spectrum in the term ``spectrum''. If
necessary we explicitly distinguish between discrete and continuous spectrum.}.

Except for the fact that they are matrices in the two-dimensional spinor space
 the operators (\ref{eq:quop}) are very similar to the fluctuation 
operators appearing in the quadratic action (\ref{eq:s2}). 
This connection becomes explicit when writing the
gauge field fluctuations in spinor space, i.e. $\slasha{a}=\sigma^\mu a_\mu$
and $\slasha{\bar a}=\sigma^\mu \bar a_\mu$. The Lagrangian density for the
gauge field fluctuations (\ref{eq:s2}) then becomes\footnote{The analogous 
relation for
the scalar fields is trivial since $\mD^2$ in (\ref{eq:quop}) is proportional to
the unit in spinor space.}:
\begin{equation}
  \label{eq:sv}
  a_\mu(\delta^{\mu\nu}\mD^2+2[\ca{F}^{\mu\nu},.\ ])a_\nu=
  \hal tr\{\slasha{\bar a}\slashed{D}\slashed{\bar D}\slasha{a}\}\quad,
\end{equation}
where $tr$ is the trace in spinor space and 
$a_\mu=\hal tr\{\bar\sigma_\mu\slasha{a}\}$. 
Later we will make use of the isospectrality of the two operators 
(\ref{eq:quop}) in computing the spectrum for the much simpler operator 
$\mD^2$ without the  additional spin-coupling. The states in question are 
then given by the non-zero eigenstates $\tilde a_\mu$  of $-\mD^2$:
\begin{equation}
  \label{eq:states}
  \slasha{a}_{(\omega)}=\frac{1}{\sqrt{\omega}}\ \slashed{D}\ \slasha{\tilde a}_{(\omega)}
\end{equation}
This will considerably simplify the 
analysis when considering less trivial self-dual 
backgrounds for non-self-dual $\theta_{\mu\nu}$. Zero-modes of the first operator
in (\ref{eq:quop}) have to be treated separately.

\subsection{Matrix models}

Above we briefly mentioned the connection between the zero-modes (\ref{eq:czm}) 
and matrix models. Since we now have discussed the fluctuation spectrum we can give 
a more detailed account on this relation. We restrict this discussion to 
self-dual $\theta_{\mu\nu}$ so that the background is supersymmetric and BPS. 
For convenience we adopt the ten dimensional language such that 
$A_M=(A_0,A_\mu,X_m)$. We are interested in the effective theory of
the $D0$-$D0$-strings  in a charge $k$ background (\ref {eq:dcart}). 
The charge $k$ background is characterised by  a projection operator $P_k$
which defines a subspace $V_k\otimes V_k^\ast$ (\ref{eq:zm}). As we have 
learned from the spectrum (\ref{eq:scsp}) the $D0$-$D0$ fluctuations live
in this subspace. According to (\ref{eq:zmspace}) and (\ref{eq:gzm}) we decompose the
ten-dimensional gauge field as follows:
\begin{equation}
  \label{eq:d0d0}
  A_M\rightarrow\ca{A}_M+a_M\quad\textrm{with}\quad 
    a_M=\sum_{i,j=0}^{k-1}(a_M)^{ij}|i\rangle\langle j|\quad.
\end{equation}
The background $\ca{A}_M$ is nontrivial only for $M=\mu$ and given by (\ref{eq:dcart}).
Analogously we expand the fermionic $D0$-$D0$ fluctuations according to:
\begin{equation}
  \label{eq:fd0d0}
  \lambda=\sum_{i,j=0}^{k-1}\lambda^{ij}|i\rangle\langle j|\quad,
\end{equation}
so that they also solve the fermionic fluctuation equation with zero 
eigenvalues, $\Gamma^MD_M\lambda=0$.

In the subspace $V_k\otimes V_k^\ast$  the connection $D_\mu$ acts trivially and the 
field strength $\ca{F}_{\mu\nu}$ is proportional to the 
unit.
Therefore one obtains by inserting  (\ref{eq:d0d0}) in the action (\ref{eq:ymrd})
the following action for the $D0$-$D0$ fluctuations:
{\setlength\arraycolsep{0pt}
\begin{eqnarray}
  \label{eq:sd0}
  &&S_{D0}=-\int\! dt\ \left [\fr{4\pi^2k}{g_{YM}^2}\right.\nonumber\\
    &&+\left.\fr{4\pi^2\theta^2}{2g_{YM}^2}\ 
           \textrm{Tr}_{U(k)}\left\{(\nabla_0\ a_I)^2-\fr{1}{2}[a_I,a_J]^2
           -2  (\lambda^T\nabla_0\lambda+\lambda^T\Gamma^I[a_I,\lambda])\right \} \right ],
\end{eqnarray}
where $\nabla_0=\partial_t+a_0$ and $I=1,\dots 9$. 
Thus up to an irrelevant additive constant,  one obtains the BFSS matrix model
\cite{Banks:1996vh} where the coupling or tension is given 
by\footnote{We follow the notation of \cite{Massar:1999jp}.} 
\begin{equation}
  \label{eq:tens}
  T_0=\sqrt{2\pi}/g=\frac{4\pi^2\theta^2}{g_{YM}^2}.
\end{equation}
It describes the dynamics of $k$ $D0$-branes by a supersymmetric 
$U(k)$ quantum mechanics obtained by the dimensional reduction of 
ten-dimensional commutative $\ca{N}=1$ $U(k)$ super Yang-Mills theory to
$0+1$ dimensions. In the $k=\infty$ limit the matrix model is conjectured 
to be equivalent to 
11-dimensional $M$-theory \cite{Banks:1996vh}. 

A different relation between matrix models and noncommutative supersymmetric 
Yang-Mills theory was given in \cite{Li:1996bi,Aoki:1999vr}. 
 The so-called  $U(N)$ IKKT matrix model 
\cite{Ishibashi:1996xs} is 
the dimensional reduction of ten-dimensional 
$\ca{N}=1$ $U(N)$ super Yang-Mills theory to zero dimensions. In the $N=\infty$ limit it 
describes noncommutative super Yang-Mills theory
by excitations around nontrivial ($D$-brane) solutions of the IKKT matrix model. 
Thus in this case the mapping between matrix model and noncommutative
super Yang-Mills theory is in the opposite way to the relation described above.

In this sense noncommutative super Yang-Mills theory plays a dual role. 
Its low energy dynamics in instanton backgrounds describing D-branes is a matrix model. 
At the same time it describes the excitations of D-brane solutions of this matrix model.

\section{Instanton backgrounds with non-self-dual $\theta_{\mu\nu}$}

We now consider  self-dual solutions in a more general $B$-field background.
In general these instantons will not be  isometric to the vacuum as was
the solution (\ref{eq:dcomp}). In complex coordinates as given above  
(\ref{eq:Az}) the self-duality equations (\ref{eq:sd1}) become 
\begin{equation}
  \label{eq:sdcom1}
  \ca{F}_{z_1\bar z_1}= \ca{F}_{z_2\bar z_2} \quad ,\quad 
  \ca{F}_{z_1\bar z_2}= 0 \quad . 
\end{equation}
The holomorphic structure of the self-duality equations becomes manifest if
one relabels the second coordinate $z_2\leftrightarrow\bar z_2$. But we prefer 
to keep the correspondence $D_{z_\alpha}\sim c^\dagger_{\alpha}$ 
for both coordinates. See also footnote \ref{fn} on page \pageref{fn}. 
Expressed through the operators (\ref{eq:fst}) the equations 
(\ref{eq:sdcom1}) read
\begin{equation}
  \label{eq:sdcom2}
  [D_{z_1},\bar D_{\bar z_1}]- [D_{z_2},\bar D_{\bar z_2}]=\fr{1}{\xi_1}-\fr{1}{\xi_2}
  \quad,\quad [D_{z_1},\bar D_{\bar z_2}]=0\quad .
\end{equation}
On the  commutative $\mathbb{R}^4$ there exists a
systematic way to construct all possible solutions to the 
self-duality equations (\ref{eq:sd1}). This is the the so-called ADHM-construction 
\cite{Atiyah:1978ri}, which can be understood by  Nahm-duality. In \cite{Nekrasov:1998ss} 
it was shown that this method can be extended to noncommutative $\mathbb{R}^4_\theta$
by adding a F(ayet)I(liopoulos)-term to the ADHM constraints. In 
\cite{Gross:2000wc} the Nahm construction of monopoles was also extended to 
 $\mathbb{R}^2_\theta\times\mathbb{R}$. In section \ref{sse:td} we observed
a duality in the spectrum of the fluxon, which is a certain limit of the monopole,
and the self-dual solution (\ref{eq:dcart}), i.e. for self-dual $\theta_{\mu\nu}$.
To make the above statements more explicit and to be able to identify 
the tachyon mode with  certain 
terms in the instanton solution for non-self-dual $\theta_{\mu\nu}$ we
briefly go through the ADHM construction of noncommutative instantons. We 
follow here the representation of \cite{Hamanaka:2003cm}.
As mentioned above, we consider the (self-dual) regime $\theta_1\geq\theta_2>0$.\\

\noindent
{\bf{ADHM construction.}} The main input for the ADHM construction
are the ADHM data. They collect the moduli of the general solution. For
a $U(N)$ $k$-instanton they are given by matrices $B_1,B_2:V\mapsto V$ and 
$I:W\mapsto V,\ J:V \mapsto W$ with the vector
spaces $V\cong\mathbb{C}^k, W\cong\mathbb{C}^N$. These matrices enter the 
ADHM constraints in the following way: Looking for a self-dual solution one
introduces the  zero-dimensional ``Dirac operator''
\begin{equation}
  \label{eq:sddir}
  \ca{D}=a + b (\sigma^\mu x_\mu\otimes\unit_k)\quad,
\end{equation}
where the $x_\mu$ are now noncommutative and the matrices $a,b$ are given by
\begin{equation}
  \label{eq:ab}
  a=\begin{pmatrix}-B_2^\dagger&B_1\\B_1^\dagger&B_2\\
      I^\dagger&J\end{pmatrix}\quad,\quad
  b=\begin{pmatrix}\unit_{2k}&0\\0&0\end{pmatrix}.
\end{equation}
The matrix $\sigma^\mu$ is given above (\ref{sdasd}). If one 
considers anti-self-dual 
solutions one has to take $\bar\sigma^\mu$ instead of $\sigma^\mu$. 
Self-duality requires
now that $\ca{D}^\dagger \ca{D}$ commutes with $\sigma^\mu$. This gives the ADHM constraints
\begin{equation}
  \label{eq:adhm}
  \mu_r:=[B_1,B_1^\dagger]+II^\dagger-J^\dagger J =\xi_2-\xi_1\quad,\quad
  \mu_c:=[B_1,B_2]+IJ=0\ \ .
\end{equation}
Here $\mu_r$, $\mu_c$ are the so called moment maps and are related to the hyper-K\"ahler
construction of manifolds. The right hand side of the $\mu_r$ equation is due 
to the noncommutativity of the coordinates $x^\mu$. This is the FI-term as it was
added in \cite{Nekrasov:1998ss}. 
For $\theta_{\mu\nu}$ being in  the self-dual regime, i.e. 
$\xi_{1,2}>0$, the FI-term vanishes at the self-dual point. This is different 
from the case considered in \cite{Nekrasov:1998ss,Nekrasov:2000zz}, 
where the duality of the $B$-field was
opposite to the duality of the gauge field, so that the FI-term is always nonzero.
The moduli $B_{1,2}$ correspond to the collective coordinates and the moduli
$I,J$  to the size of the (multi) instantons. If the group is U(1) the self-dual point $\xi_1=\xi_2$ 
corresponds to  the small instanton singularity of the moduli space. 
For generic $B$-field, i.e. for  $\xi_1\neq\xi_2$, this 
singularity is resolved.

Next one has to solve the zero-dimensional ``Dirac equation'' and conditions  
\begin{equation}
  \label{eq:sddir2}
  \ca{D}^\dagger\Psi=0\quad,\quad \Psi^\dagger\Psi=1\quad,\quad
  \Psi\Psi^\dagger +\ca{D}\fr{1}{\ca{D}^\dagger\ca{D}}\ca{D}^\dagger=\unit\quad.
\end{equation}
The self-dual connection is then given by 
\begin{equation}
  \label{eq:sdcon}
  D_\mu=i\Psi^\dagger\theta^{-1}_{\mu\nu}x^\nu\Psi\quad.
\end{equation}
The second relation in (\ref{eq:sddir2}) is just a normalisation condition, where 
the last equation is  a completeness relation, which is trivially fulfilled in 
commutative space but not in noncommutative space 
\cite{Chu:2001cx,Lechtenfeld:2001ie,Ivanova:2004cq,Tian:2002si,Correa:2001wv}\footnote{We 
are grateful to Kirsten Vogeler for pointing this out to us.}. 
This condition will be crucial for the interpretation of our solution.\\

\noindent
{\bf{Nahm- and $T$-duality.}}
The appearance of the zero-dimensional ``Dirac equation'' in the above construction
has its origin in  Nahm-duality. The Nahm transformation maps the moduli space
for $U(N)$ $k$-instantons on the four-torus $T^4$ to the moduli space
of $U(k)$  $N$-instantons on the dual torus $\hat T^4$:
\begin{equation}
  \ca{M}_{ T^4}^{k,N}\simeq\ca{M}_{\hat T^4}^{N,k}\quad.
\end{equation}
By sending the dual radii to zero
the original four torus approaches $\mathbb{R}^4$ or $\mathbb{R}^4_\theta$. 
Thus the dual equations are 
zero-dimensional and therefore  can be solved for particular cases. 
The original connection is then  given by the 
Nahm transformation (\ref{eq:sdcon}). On the string theory side the corresponding
duality is $T$-duality. It maps the moduli space of $k$ $D0$-branes sitting on a stack
 of $N$ $D4$-branes wrapped around a four-torus $T^4$ to the moduli space of
 $N$ $D0$-branes on  $k$ $D4$-branes wrapped around the dual  torus $\hat T^4$.
 
A similar construction, the so called Nahm construction, 
exists also for monopoles on $\mathbb{R}^3$ or 
$\mathbb{R}^2_\theta\times\mathbb{R}$ \cite{Nahm:1979yw,Craigie:1982gq,Gross:2000wc}. 
For this one shrinks
one radius of $T^4$ to zero and sends the other ones to infinity. The dual equations
are thus one-dimensional (ordinary) differential equations. The Nahm transformation 
then gives the monopole gauge field and the nontrivial Higgs field.
The corresponding brane picture for a $U(2)$ monopole is a $D1$-brane/string 
stretched 
between two separated $D3$-branes \cite{Diaconescu:1996rk}. The fluxon is obtained
by sending one of the $D3$-branes to infinity. Thus the distance
between them 
and  therefore the mass of the
massive fields become infinite. The massive fields decouple and only the 
massless $U(1)$ fields remain \cite{Gross:2000ph}.   

Now $T$-duality is a more general concept than  Nahm transformations tailored for
self-dual connections. As discussed before, by $T$-dualising along the 
$D$-string which pierces the
 remaining $D3$-brane with a rank two $B$-field on it  one 
obtains a $D0$-$D4$ system with a rank four $B$-field. 
$T$-duality also acts on the string fluctuations
on the branes. So in principle one should find this duality also for the quantum 
fluctuations of the low-energy effective descriptions of these brane systems, 
at least 
for that part of the spectrum that survives the zero-slope limit.
Therefore we think that the equivalence of the fluctuation spectra 
of the fluxon and the instanton for self-dual $\theta_{\mu\nu}$ that we found before
is due to the $T$-duality in the $D$-brane description. The classical action/energy 
of the two objects is different but this is not surprising, since $T$-duality 
changes 
the charges of $D$-branes. As mentioned above we have to leave a  more detailed description
for future studies.\\

\noindent
{\bf{The single $U(1)$ instanton.}} In the following we concentrate on the 
case\footnote{For noncommutative $U(2)$ multi-instanton configurations see e.g.
\cite{Chu:2001cx,Furuuchi:2000dx,Lechtenfeld:2001ie,Horvath:2002bj,Ivanova:2004cq}.}  
$k=N=1$. For 
simplicity we also set the translational moduli $B_{1,2}$ in (\ref{eq:adhm}) to zero.
So for $\xi_1>\xi_2$ 
\begin{equation}
  \label{eq:size}
  |J|^2=\xi_1-\xi_2\quad,\quad I=0
\end{equation}
solves (\ref{eq:adhm}). Now we can see that $|J|=\sqrt{\xi_1-\xi_2}$ is the size 
of the 
instanton and therefore the instanton shrinks to 
zero  at the self-dual point\footnote{The notion 
of ``size''  has to be taken with care in noncommutative space.}.   
Following the steps  described in (\ref{eq:sddir2}), (\ref{eq:sdcon}) one obtains the
following solution (from now on we use complex coordinates if not 
stated differently and we abbreviate $D_{z_\alpha}$ with $D_\alpha$):
\begin{eqnarray}
  \label{eq:sol}
  D_1&=&\fr{1}{\sqrt{\xi_1}}S\Lambda^\halb\ c_1^\dagger\ \Lambda^{-\halb}S^\dagger
  \quad,\nonumber\\
  D_2&=&\fr{1}{\sqrt{\xi_2}}S\Lambda^{-\halb}\ c_2^\dagger\ \Lambda^{\halb}S^\dagger
    +\fr{J}{\sqrt{\xi_1\xi_2}}SP\quad,
\end{eqnarray}
where the shift operator $S=S^{\textrm{II}}$ and the projection operator 
$P$ are defined  in (\ref{eq:shift}) and (\ref{eq:P}). The operator 
$\Lambda$ is given by 
\begin{equation}
  \label{eq:Lam}
  \Lambda=\frac{M+\xi_1}{M+\xi_2}\quad,\quad M=\xi_1 n_1 +\xi_2 n_2\quad,
\end{equation}
where $n_\alpha$ are the number operators of the two Hilbert spaces
(\ref{eq:fock}). The adjoint connections are given by 
$\bar D_\alpha=-D_\alpha^\dagger$. The solution (\ref{eq:sol}) is by 
construction a self-dual charge one field configuration. Thus its five brane mass 
according to (\ref{eq:top}) is given by
\begin{equation}
  \label{eq:m5}
  M_{5}^{BPS}=\frac{4\pi^2}{g_{YM}^2}\quad.
\end{equation}

The instanton (\ref{eq:sol}) differs in two respects
from the instanton at the self-dual point $\xi_1=\xi_2$ (\ref{eq:dcomp}). First there 
is an additional ``axial'' term proportional to the size $J$  of the instanton. This
term is crucial for the completeness relation in (\ref{eq:sddir2}). 
As we will see below 
it  has its  origin in the tachyonic mode that we found for the unstable 
solitonic state (\ref{eq:tach1}). 
Second
the creation/annihilation operators are dressed by the operators $\Lambda$, which 
``entangle'' the two Hilbert spaces. In the self-dual limit $\xi_2\rightarrow\xi_1$
the solution (\ref{eq:sol}) smoothly approaches the solution (\ref{eq:dcomp}) 
at the self-dual point.

\subsection{Fluctuation spectrum}

As we  show in the next section  
the instanton (\ref{eq:sol}) will be the final state of
the tachyon condensation process of the unstable configuration (\ref{eq:dcart}). To
characterise this final state in more detail we investigate the fluctuation spectrum, 
i.e. the asymptotic quantum states in the presence of  the background (\ref{eq:sol}).
The background is now much more complicated than before and it does not seem  
possible to solve the fluctuation equation for the  gauge field modes. But as we have 
shown above, also in the noncommutative case the gauge-field fluctuation operator
is, except for zero-modes, isospectral to the much simpler operator $\mD^2$ 
(\ref{eq:quop}).  Possible zero-modes will be treated separately. Thus we are looking 
for solutions of  
\begin{equation}
  \label{eq:eveq}
  -\Delta\vf_\omega=\omega^2\vf_\omega\quad\textrm{with}\quad
  \Delta=2\sum_{\alpha=1,2}\left([D_\alpha,[\bar D_\alpha,.\ ]]
  +[\bar D_\alpha,[D_\alpha,.\ ]]\right)\ \ ,
\end{equation}
where $D_\alpha$ is given by (\ref{eq:sol}).
Because of the axial term in (\ref{eq:sol}) there are no longer any finite-dimensional 
invariant subspaces. Nevertheless it will prove to be useful to
decompose the operator $\vf_\omega$ as before  according to the 
sectors $V_{00},V_{04,40}$ and $V_{44}$ (\ref{eq:fl1}). Let 
us now specify  the different terms in (\ref{eq:fl1}) as follows:
\begin{eqnarray}
  \label{eq:abd}
  a&=&\sum_{m,n}a_{mn}|0,0\rangle\langle m,n|\quad,\quad 
  b=\sum_{m,n}b_{mn}|m,n\rangle\langle 0,0|\quad,\nonumber\\
  d&=&\sum_{m_\alpha,n_\alpha} d^{\ m_1,m_2}_{\ n_1,n_2}\ |m_1,m_2\rangle\langle n_1,n_2|
  \quad,\quad \alpha\in\mathbb{C}\quad.
\end{eqnarray}
The index $\alpha$  labels the two noncommutative planes described by 
the two parameters $\xi_\alpha=2\theta_\alpha$ (not to be confused with 
the coefficient $\alpha\in\mathbb{C}$ in (\ref{eq:fl1})). 
If not stated differently summations always 
run from zero to infinity. 
Introducing the abbreviations
\begin{eqnarray}
  \label{eq:abv}
  &&Z_1^\dagger:=\Lambda^\halb\ c_1^\dagger\ \Lambda^{-\halb}
  =\sum_{p,q}A_{p,q}|p+1,q\rangle\langle p,q|\quad,\nonumber\\
  &&Z_2^\dagger:=\Lambda^{-\halb}\ c_2^\dagger\ \Lambda^{\halb}
  =\sum_{p,q}B_{p,q}|p,q+1\rangle\langle p,q|
\end{eqnarray}
for the dressed creation operators and $\rho:=\fr{J}{\sqrt{\xi_1\xi_2}}$ for 
the ``strength''  of the axial part in 
(\ref{eq:sol}) the fluctuation equation (\ref{eq:eveq}) in the different sectors are 
of the form:
{\setlength\arraycolsep{2pt}
\begin{eqnarray}
  \label{eq:sec}
  V_{00}&:&\quad \alpha(4|\rho|^2-\omega^2)-4|\rho|^2d_{0,0}^{0,0}=0\ \ ,\nonumber\\
  V_{04}&:&\quad a_{mn}(2y|\rho|^2-\omega^2+2N_{mn})
                      -\frac{4\bar \rho}{\sqrt{\xi_2}}B_{mn}d_{m,n+1}^{0,0}=0\ \ ,
                      \nonumber\\
  %a_{00}(4|t|^2-\omega^2+2N_{00})-\frac{4\bar t}{\sqrt{\xi_2}}d_{0,1}^{0,0}=0\nonummber\\
  V_{40}&:&\quad b_{mn}(2y|\rho|^2-\omega^2+2N_{mn})
                      -\frac{4 \rho}{\sqrt{\xi_2}}B_{mn}d^{m,n+1}_{0,0}=0\ \ ,
  % b_{00}(4|t|^2-\omega^2+2N_{0,0})-\frac{4 t}{\sqrt{\xi_2}}d^{0,1}_{0,0}=0\nonummber\\
\end{eqnarray}}
where $y=2$ for $\{m,n\}=\{0,0\}$ and equal to one otherwise. We also introduced the 
abbreviation
\begin{equation}
  \label{eq:Nmn}
  N_{mn}:=\fr{1}{\xi_1}(A_{m-1,n}^2+A_{mn}^2)+
      \fr{1}{\xi_2}(B_{m,n-1}^2+B_{mn}^2)\quad.
\end{equation}

 The $V_{44}$-sector equation is
more involved\footnote{Here and in the following, terms with negative indices 
are not present. 
Equivalently they can be put to zero whenever such factors formally  
appear in an equation.}:
{\setlength\arraycolsep{0pt}
\begin{eqnarray}
  \label{eq:44}
 && 2\sum_{\alpha=1,2}\frac{1}{\xi_\alpha}\left([Z_\alpha,[Z_\alpha^\dagger,d]]+
    [Z_\alpha^\dagger,[Z_\alpha,d]]\right)-\omega^2 d   \nonumber\\
  && -\fr{4\rho}{\sqrt{\xi_2}}\sum_{m,n} a_{m,n-1}B_{m,n-1}|00\rangle\langle mn|
     -\fr{4\bar \rho}{\sqrt{\xi_2}}\sum_{m,n} b_{m,n-1}B_{m,n-1}|mn\rangle\langle 00|
     \nonumber\\
  &&-4|\rho|^2\alpha|00\rangle\langle 00|
     +2|\rho|^2\sum_{m,n}\left(d_{mn}^{0,0}|00\rangle\langle mn|
       +d_{0,0}^{mn}|mn\rangle\langle 00|\right)=0.
\end{eqnarray}}
By inspection of the equations (\ref{eq:sec}), (\ref{eq:44}) one can see that the axial
term in the connection (\ref{eq:sol}) introduces a mixing of the different 
sectors in 
a very special way. 
The equations (\ref{eq:sec}) fix  the $V_{44}$ operator $d$ 
at the ``boundaries'' of
the two noncommutative planes, whose Hilbert space descriptions are 
parametrised by two 
lattices $\{m_\alpha,n_\alpha\}$ with
$m_\alpha,n_\alpha\geq0$. Thus the solution for the eigenvalue equation 
is obtained by 
implementing  these boundary conditions on the general solution of 
the $d$-\emph{equation}
\begin{equation}
  \label{eq:d}
  H\ d=\epsilon^2\ d\quad,
\end{equation}
where this is essentially the first line in (\ref{eq:44}) which also defines 
the operator $H$.
The boundary conditions then fix the mode energies $\omega^2$. 
Let us assume for the moment that the coefficients of $\alpha$, $a_{mn}$ 
and $b_{mn}$ 
in (\ref{eq:sec}) do not vanish. We will comment on the opposite case below. 
Then using
(\ref{eq:d}) equation (\ref{eq:44}) can be written as
{\setlength\arraycolsep{0pt}
\begin{eqnarray}
  \label{eq:d44}
  &&(\epsilon^2-\omega^2)d=\nonumber\\
  &&\frac{16|\rho|^2}{\xi_2}\sum_{m,n}\frac{B_{m,n-1}^2}{2y|\rho|^2-\omega^2+2N_{m,n-1}}
  \left( d_{m,n}^{00}|00\rangle\langle mn|+d^{m,n}_{00}|mn\rangle\langle 00|\right)
   \nonumber\\ 
 && +\frac{16|\rho|^4}{4|\rho|^2-\omega^2}\ d_{0,0}^{0,0}\ P -
  2|\rho|^2\sum_{m,n}\left(d_{mn}^{0,0}|00\rangle\langle mn|
    +d_{00}^{mn}|mn\rangle\langle 0,0|\right) .
\end{eqnarray}}
 Thus solving  the problem (\ref{eq:d}) is crucial to get the fluctuation spectrum.\\

\subsection{The disentangled system} 

The operators (\ref{eq:abv}) have the same mapping property as the
undressed creation operators in the sense that they leave the following 
subspaces of $\ca{H}\otimes \ca{H}^\ast$ invariant:
\begin{equation}
  \label{eq:online}
  \Psi_{a,b}=\sum_{m,n}O_{mn}|m+a,n+b\rangle\langle mn|\quad, 
\end{equation}
and analogously for other combinations of shifts in the occupation number. 
To understand the general structure of the (quantum) states 
in the presence of the instanton background we discuss a simplified 
version of (\ref{eq:d}). Let us ``switch'' off the entanglement by setting the 
operators $\Lambda$ to the unit in (\ref{eq:sol}). Thus equation (\ref{eq:d}) becomes
the free equation (\ref{v44}). But the noncommutative plane waves 
(\ref{eq:ncpw}) are an improper basis of solutions in this case. 
The boundary conditions imposed 
by (\ref{eq:sec}) or (\ref{eq:d44}), respectively, overconstrain the system since
the noncommutative plane wave does not vanish on the whole boundary, except for zero
momentum. The appropriate basis to fulfil these boundary conditions is of the form 
(\ref{eq:online}). Operators of the form (\ref{eq:online}) are non-vanishing only 
for one point at the boundary 
of the lattice $m_\alpha,n_\alpha\geq0$, thus giving one condition to 
determine the single free 
parameter $\omega^2$. For the simplified case $\Lambda=\unit$ the equation (\ref{eq:d}) 
can  easily be Weyl-transformed to the star-product representation by the inverse version
of (\ref{eq:wmap}). Because of the vanishing gauge field 
for $\Lambda=\unit$,   $H$  simply becomes the four-dimensional Laplacian 
in the space of ordinary functions. \\

\noindent
{\bf{Discrete spectrum.}} The ansatz (\ref{eq:online}) is realized
by polar coordinates $r_\alpha,\vf_\alpha$ for the two noncommutative planes. 
Therefore the general solution of (\ref{eq:d}) according to the ansatz 
(\ref{eq:online})
in the inverse Weyl-transformed version is given by
\begin{equation}
  \label{eq:bessel}
  \fr{1}{2}\ e^{ia\vf_1}J_a(k_1r_1)\ e^{ib\vf_2}J_b(k_2r_2)\quad,\quad 
  \epsilon^2=k_1^2+k_2^2\quad,
\end{equation}
where $J_p$ is the order $p$ Bessel function of the first kind \cite{Leb:SF}, 
which are regular at the origin, 
and $k_1^2=k_{x_1}^2+k_{x_2}^2$,  $k_2^2=k_{x_3}^2+k_{x_4}^2$ are the total 
momenta in 
the two planes. 
Normalisability of the states implies $k_{1,2}>0$ and thus the spectrum is positive. 
The eigenstates  according to the ansatz 
(\ref{eq:online}) are angular momentum eigenstates, with angular momentum quantum numbers 
$a,b$ in the first/second noncommutative plane.  
Weyl-transforming (\ref{eq:bessel}) back to the operator representation one obtains the 
coefficients in (\ref{eq:online}) as follows:
\begin{eqnarray}
  \label{eq:coef}
  O_{mn}(a,b,k)=\sqrt{\fr{m!n!}{(m+a)!(n+b)!}}\left(\fr{k_1\theta_1}{2}\right)^{a/2}
    \left(\fr{k_2\theta_2}{2}\right)^{b/2}\times \nonumber \\ 
    e^{-\frac{k_1^2\theta_1+k_2^2\theta_2}{4}}\ 
      L_{m}^a(\fr{k_1^2\theta_1}{2})L_n^b(\fr{k_2^2\theta_2}{2})\quad,
\end{eqnarray}  
where the $L_m^a$ are the Laguerre polynomials \cite{Leb:SF}. As one can see, the boundary 
matrix element $O_{0,0}$ vanishes only for  infinite momentum in one of the two planes,
so that for asymptotic high momentum the states would no longer feel the instanton 
background. But for generic momentum one obtains  from 
(\ref{eq:d44}) the following conditions: The ``bulk'' matrix elements imply 
$\epsilon^2=k^2=\omega^2$ whereas from the boundary matrix element one obtains
\begin{equation}
  \label{eq:eigenval}
  \omega^2_{ab}=2\left(y|\rho|^2+N_{a,b-1}-\fr{4}{\xi_2}\ B_{a,b-1}^2\right)\quad 
  \textrm{for }a,b\neq 0
%  \nonumber\\   \omega^2_{0,0}&=&0\quad,
\end{equation}
where $y=2$ for $a=0,b=1$ and equal to one otherwise. The diagonal operator
with $a,b=0$ would be a zero-mode, i.e. $\omega^2_{0,0}=0$,   
but this mode is not normalisable and therefore has  
to be excluded.  By hermiticity one obtains the same frequencies for the hermitian 
conjugate of the operator (\ref{eq:online}). 

For the disentangled case, where $\Lambda=\unit$, the elements $N_{mn}, B_{mn}$
simplify to 
\begin{eqnarray}
  \label{eq:nb}
  N_{mn}=\fr{2m+1}{\xi_1}+\fr{2n+1}{\xi_2}\quad,\quad B_{mn}=\sqrt{n+1}\quad,
\end{eqnarray}
so that with $|\rho|^2=(1/\xi_2-1/\xi_1)$ we obtain for the mode energies
\begin{eqnarray}
  \label{eq:mfree}
  \omega^2_{ab}= 4(\frac{a}{\xi_1}-\frac{b}{\xi_2})\quad \textrm{with } b>0\quad.
\end{eqnarray}
The mode $\Psi_{0,1}$ is negative, i.e $\omega^2_{0,1}=-2(1/\xi_1+1/\xi_2)<0$, and 
 therefore is also non-normalisable. But also for other values of $a,b$  normalisability 
excludes some of the modes with energy (\ref{eq:mfree}). Positivity of the 
mode energies (\ref{eq:mfree}) requires 
\begin{equation}
  \label{eq:pos}
  a>\fr{\xi_1}{\xi_2}\ b\quad,
\end{equation}
which, depending on the relative size of the noncommutative parameters, excludes a number 
of additional modes from the discrete spectrum. Note that  
we always have $\xi_1>\xi_2$. \\

\noindent
{\bf{Continuous spectrum.}} 
In addition to the ansatz (\ref{eq:online}) one can find compatibility with the boundary 
terms in (\ref{eq:44}) for operators of the following form:
\begin{equation}
  \label{eq:online2}
  \Psi_{ab}=\sum_{m}O_{mn}|m+a,n\rangle\langle m,n+b|\quad\textrm{with } a,b\neq 0.
\end{equation}
For this operator all boundary matrix elements  that occur in the equations 
(\ref{eq:sec}), (\ref{eq:44}) are zero. Thus  a solution of (\ref{eq:d}) 
is automatically a solution of the whole system with $\alpha=a_{mn}=b_{mn}=0$ if 
$\epsilon^2=\omega^2$.
It turns out that the coefficients $O_{mn}$ in (\ref{eq:online2}) are also given 
by (\ref{eq:coef}). The eigenvalues are again given by 
\begin{equation}
  \label{eq:con}
  \epsilon^2=\omega^2=k_1^2+k_2^2\quad,
\end{equation}
where we again use the notation as introduced in (\ref{eq:bessel}), but the 
momenta are now unconstrained except for the positivity condition $k_\alpha>0$. 
Thus one also finds a continuous spectrum.

It can be easily seen, that if the coefficients of $\alpha,a_{mn},b_{mn}$ in 
(\ref{eq:sec}) vanish 
one does not obtain new solutions.

\subsection{Entangled system}

Let us now discuss how the above results will be modified if one takes the nontrivial
$\Lambda$ (\ref{eq:Lam}) into account. The  equations 
(\ref{eq:sec}), (\ref{eq:44}) as well as the invariant subspaces 
(\ref{eq:online}), (\ref{eq:online2}) are unchanged, but the 
matrix elements $B_{mn}$ and $N_{mn}$ are less trivial now. Therefore also  
equation (\ref{eq:d}) becomes more complicated and 
the mode-functions will  not be of the simple
form (\ref{eq:bessel}) but will include nontrivial  form factors, such that  
the solutions no longer factorise according to the two noncommutative planes.
In operator language this means that the matrix elements $O_{mn}$ 
 will include these form factors and will also no longer factorise. So
 $\Psi_{ab}$ can no longer be written as a sum of  products of two operators. Therefore 
we use the notion (dis)entangled. But since the general structure is unchanged 
  the spectrum again separates into a discrete part and 
a gap-less continuum as for the disentangled case. 
The relation for the discrete spectrum 
(\ref{eq:eigenval}) also remains valid, 
but now one has to use in (\ref{eq:Nmn}), (\ref{eq:eigenval})  
the nontrivial matrix elements  
\begin{equation}
  \label{eq:nb2}
   A_{mn}^2=(m+1)\frac{\Lambda_{m+1,n}}{\Lambda_{mn}}\quad,\quad 
   B_{mn}^2=(n+1)\frac{\Lambda_{mn}}{\Lambda_{m,n+1}}\quad ,
\end{equation}
where  the matrix elements $\Lambda_{mn}$ can be read off (\ref{eq:Lam}). 
The mode energies for the states $\Psi_{ab}$ are now given by ($b>0$):
\begin{equation}
  \label{eq:fmode}
  \omega^2_{ab}=4\left(\frac{a}{\xi_1}-\frac{b}{\xi_2}+\frac{a+b-1}{a\xi_1+(b-1)\xi_2}
    -\frac{2(a+b)}{a\xi_1+\xi_2}+\frac{a+b+1}{(a+1)\xi_1+b\xi_2}\right)
\end{equation}
As one can see the, mode energies (\ref{eq:eigenval}) are now
rather complicated functions of $a,b$, but they describe the  exact spectrum in the 
background (\ref{eq:sol}). Again one has to check if there are certain
values $a,b$ which have to be excluded from the spectrum. This time we do not
have explicit solutions for the operators $\Psi_{ab}$ so that we cannot 
explicitly exclude negative modes by normalisation conditions. But 
we expect that this property will not be changed by nontrivial form factors.
Anyway, from the supersymmetry or factorisation property  (\ref{eq:quop})
of the fluctuation operators with self-dual classical background fields one knows that the 
spectrum has to be positive. Thus as before we exclude modes with $\omega^2_{ab}<0$
from the spectrum. It is easy to see that $\omega^2_{ab}>0$ if 
\begin{equation}
  \label{eq:neg}
  a > \frac{\xi_1}{\xi_2}\  b+ \epsilon\quad,
\end{equation}
where $\epsilon<1$. Thus one obtains roughly the same condition  as for the
disentangled system. The diagonal operator $\Psi_{00}$ would be a zero-mode, but 
as before 
we expect that this mode as the zero momentum threshold mode of the continuum 
is not normalisable, or equivalently, the field strength $\ca{F}_{\mu\nu}$ is no longer
 reducible. We will comment on this in the discussion below.\\

\noindent
{\bf{Zero-modes.}} So far we have investigated the operator $-\mD^2$ which is except
for zero-modes isospectral to the operator in question (\ref{eq:quop}). Since 
this time  $-\mD^2$ has no zero-modes the kernel of the nontrivial
operator in (\ref{eq:quop}) and the kernel of $\slashed{\bar D}$ coincide:
\begin{equation}
  \label{eq:kern}
  \textrm{kern}\{\slashed{\bar D}\}=\textrm{kern}\{\slashed{D}\slashed{\bar D}\}\quad.
\end{equation}
But the number of bosonic zero-modes  is twice the number
of single solutions to $\slashed{\bar D}\chi= 0$ \cite{Belitsky:2000ws}. Since 
now $-\mD^2$ is strictly positive the considerations of \cite{Kim:2002qm} are valid
without any restrictions. Thus for a $U(N)$ $k$-instanton one obtains $4Nk$ 
bosonic zero-modes. In our case these  are four zero-modes 
corresponding to translations of the $D0$-brane inside the $D4$-brane, or more
exactly, translations of the $D0$-$D4$-bound state object. Since for vanishing 
scalar fields 
$\slashed{\bar D}\chi= 0$ is a fermionic equation of motion in the 
Euclidean four-dimensional theory a rather simple method to obtain solutions 
to it is via the broken supersymmetries in the instanton background. Thus 
one obtains the fermionic zero-modes as
\begin{equation}
  \label{eq:fzm}
  \chi_\pm\sim \sigma^{\mu\nu}\ca{F}_{\mu\nu}\epsilon_\pm\quad,
\end{equation}
where $\epsilon_\pm$ are two linearly independent two component spinors. 
Hence the number of bosonic zero-modes is indeed four, i.e. $4Nk$ with $N=k=1$. 

\subsection{Discussion of the spectrum}

As a first difference to the spectrum for the small instanton, i.e. for self-dual 
$\theta_{\mu\nu}$, or the unstable state with non-self-dual $\theta_{\mu\nu}$, 
respectively
we note that the operator $-\mD^2$ has now no zero-modes. As we discussed above
(\ref{eq:zm1}) these zero-modes correspond to translational moduli of the $D0$-brane(s)
transverse to the $D4$-brane. Thus for the instanton solution away from the 
self-dual point 
(\ref{eq:sol}) the $D0$-branes are confined in the $D4$-brane. This confirms the 
description as given in \cite{Furuuchi:2000dx} for example. For a self-dual 
$B$-field the lower dimensional brane of a $Dp-D(p+4)$ system can freely move 
transverse to the higher dimensional brane and thus separate from the $D(p+4)$-brane. The 
effective theory on the lower dimensional brane is in the Coulomb phase. On the contrary
for non-self-dual $\theta_{\mu\nu}$ there is no small instanton singularity and the 
lower dimensional brane is confined in the $D(p+4)$-brane. In this case the effective 
theory on the lower dimensional brane gains a FI-term and is thus in the Higgs
phase. This is  reflected by the axial term in the solution (\ref{eq:sol}) living 
 on the higher dimensional brane. This axial- or FI-term is  
also related to the appearance of 
the tachyonic mode (\ref{eq:tach1}) for the unstable configuration as we will see.
Because of the mixing property of the axial term in the Hilbert space the 
identification of the $D0$-brane according to $(0,0)$, $(0,4)$ and $(4,4)$
fluctuations is no longer possible. In forming a $D0$-$D4$ bound state the $D0$-brane
looses its strict individual character. 

Second, half of the continuum states are  tied up with the states of the 
discrete spectrum, such that 
the number or density of continuous states is reduced. Because of the 
condition (\ref{eq:neg}) the number of discrete states is also reduced compared to
the self-dual point. The eigenvalues for the discrete states (\ref{eq:fmode}) are
exact.

The continuous and discrete states are angular momentum eigenstates w.r.t. 
the angular momenta in the two noncommutative planes. States with the same orientation
of the angular momenta in the two planes (\ref{eq:online}) are discrete states, whereas 
states with opposite angular momentum orientations in the two planes (\ref{eq:online2}) 
remain continuum states.
 
As expected the number of  zero-modes in the gauge field fluctuations is
unchanged. They  describe the translational moduli of the $D0$-$D4$-bound state object.   

One set of possible additional discrete states could not be investigated here. For 
entangled system, i.e. $\Lambda\neq \unit$, the equation (\ref{eq:d}) could have 
additional bound states. To find them one would have to solve (\ref{eq:d}) explicitly, 
which is a considerably hard problem.

\section{Tachyon condensation}

In the last section we have constructed a (self-dual) BPS state (\ref{eq:sol})
for non-self-dual $\theta_{\mu\nu}$ which takes values in the self-dual regime
$\theta_1>\theta_2>0$. We characterised this state also by its excitation spectrum. 
In this section we will argue that this state is the final state 
of the tachyon condensation process of the unstable $D0$-$D4$ system (\ref{eq:dcomp}) 
which was described above. A first hint to the fact  
that the unstable state  decays into
the BPS state (\ref{eq:sol}) and not to the vacuum, like 
unstable solitonic states in $2+1$ dimensions \cite{Aganagic:2000mh}, 
is given by energetic considerations. The BPS state (\ref{eq:sol}) is a stable state and 
its energy (mass) (\ref{eq:m5}) is below the energy of the unstable state (\ref{eq:top}).
But we will give more convincing arguments below.

\subsection{Relevant degrees of freedom}  

Let us  first write the tachyonic mode in (\ref{eq:04gf}) 
in complex coordinates. The total tachyonic excitation of the gauge field 
is an anti-hermitian  linear combination of the $V_{04}$ and $V_{40}$ 
sector tachyonic fluctuations. This way one obtains
\begin{equation}
  \label{eq:tm}
  a^{th}_{z_2}=t\ SP\quad,\quad\bar a^{th}_{\bar z_2}=-\bar t\ PS^\dagger\quad,\quad
  a^{th}_{z_1}=\bar a^{th}_{\bar z_1}=0\quad,
\end{equation}
where $t$ is the (complex) tachyon amplitude. Thus exciting the tachyonic mode
on the unstable background (\ref{eq:dcomp}) adds an operator of the form 
$|0,1\rangle\langle 0,0|$ to $D_2$. The idea for calculating the tachyon potential 
is to neglect the exact time evolution of the condensation process 
and to consider only the potential for static fields. The kinetic term remains unknown. 
By ``integrating out'' all degrees of freedom
except the tachyon mode one obtains the potential as a function of the tachyon 
amplitude only, which is then the tachyon potential. 
Here ``integrating out'' means minimising the potential and inserting the 
solutions for the relevant degrees of freedom. Since we consider now only 
 static configurations it is convenient to work in the temporal gauge $A_0=0$.
For static configurations the associated Gau{\ss}  law constraint is automatically 
fulfilled.

The initial state, i.e. the unstable solution (\ref{eq:dcomp}) including the 
tachyonic excitation (\ref{eq:tm}) as well as the final state (\ref{eq:sol})
are  operators of the form $~ \sum_{p,q} L_{pq}|p+1,q\rangle\langle p,q|$ or 
with shifted arguments in the second Hilbert space factor and of course the 
hermitian conjugates thereof. Thus following the symmetry arguments of  
\cite{Aganagic:2000mh} we make the following ansatz for the relevant degrees of
freedom\footnote{In the following we suppress 
the dependence on the complex conjugate amplitude $\bar t$ in our notation. 
%The operators which are considered here are \emph{not} holomorphic in the tachyon amplitude.
}:
\begin{eqnarray}
  \label{eq:ans}
  \nabla_1=\fr{1}{\xi_1}SZ_1^\dagger(t)S^\dagger\quad,\quad
  \nabla_2=\fr{1}{\xi_2}SZ_2^\dagger(t)S^\dagger+tSP\quad,
\end{eqnarray}
and $\bar\nabla_\alpha=-(\nabla_\alpha)^\dagger$. The operators $Z_\alpha^\dagger(t)$
are of the form 
\begin{equation}
  \label{eq:Z}
  Z_1^\dagger(t)=\sum_{p,q}A_{p,q}(t)|p+1,q\rangle\langle p,q|\ ,\
  Z_2^\dagger(t)=\sum_{p,q}B_{p,q}(t)|p,q+1\rangle\langle p,q|\ .
\end{equation}
The operators (\ref{eq:Z}) should not be confused with the operators (\ref{eq:abv}) 
but we will see that after the condensation process they will coincide.
The ansatz (\ref{eq:ans}), (\ref{eq:Z}) implies that not only the initial and final state
are of this form, but also during the condensation process only fields of this form
will contribute. On this reasoning also the first positive excitation in 
(\ref{eq:04gf}) should have been included. It would add a 
term to the operator $\nabla_1$ which is  analogous
to the tachyon mode. But as it turns out by 
integrating out the non-tachyonic degrees of freedom the amplitude of this mode
vanishes. Putting the scalar fields and  fermions to zero also satisfies the
minimum condition for the potential. 

Thus, inserting the ansatz (\ref{eq:ans}) 
into the action (\ref{eq:ymrd}), one obtains the following potential\footnote{Again we 
omit the 
time integral from the definition  (\ref{eq:tr}).}
{\setlength\arraycolsep{0pt}
\begin{eqnarray}
  \label{eq:pot1}
  V=\frac{2}{g_{YM}^2}\int \textrm{Tr}&&
  \left [\sum_{\alpha}([\nabla_\alpha,\bar\nabla_\alpha]-\fr{1}{\xi_\alpha})^2\right .
    \nonumber\\
    &&\left . -\{[\nabla_1,\nabla_2],[\bar\nabla_1,\bar\nabla_2]\}
    -\{[\nabla_1,\bar\nabla_2],[\bar\nabla_1,\nabla_2]\}\right]\  .
\end{eqnarray}}
Minimising the potential (\ref{eq:pot1}) w.r.t. 
$Z_\alpha(t)$ one obtains (suppressing the $t$-dependence in our notation)
 the following equations:
\begin{eqnarray}
  \label{eq:eomZ}
  [Z_1^\dagger,\ca{K}]-|t|^2 Z_1^\dagger P+\fr{2}{\xi_2}[Z_2^\dagger,[Z_1^\dagger,Z_2]]&=&0
 \ \ ,\nonumber\\
  {[Z_2^\dagger,\ca{K}]}-|t|^2 Z_2^\dagger P+\fr{2}{\xi_1}[Z_1^\dagger,[Z_2^\dagger,Z_1]]&=&0
 \ \ ,
\end{eqnarray}
where we have introduced the operator 
\begin{equation}
  \label{eq:K}
  \ca{K}=\fr{1}{\xi_1}[Z_1^\dagger,Z_1]-\fr{1}{\xi_2}[Z_2^\dagger,Z_2]\quad,
\end{equation}
and we have used that $Z_\alpha P=0$ due to the  ansatz (\ref{eq:Z}).
To find an exact solution to the nonlinear equations (\ref{eq:eomZ}) is rather 
difficult. Note that in the 2+1 dimensional case, as it was considered in 
\cite{Aganagic:2000mh}, the
analogous equations become linear. The initial and final state 
fall into a certain class of solutions to (\ref{eq:eomZ}) which are of the form 
\begin{equation}
  \label{eq:eomZ2}
  \ca{K}(t)=\fr{|J|^2}{\xi_1\xi_2}+|t|^2 P\quad,\quad [Z_1^\dagger,Z_2]=0\quad.
\end{equation}
We can an only give an approximate solution to (\ref{eq:eomZ}):
\begin{eqnarray}
  \label{eq:way}
  \langle pq|Z_1^\dagger|pq\rangle &=&A_{pq}=
  \sqrt{p+1}\sqrt{1-|\tau|^2\left(1-\frac{\Lambda_{p+1,q}}{\Lambda_{pq}}\right)}
  \nonumber\\
  \langle pq|Z_2^\dagger|pq\rangle &=&B_{pq}=
  \sqrt{q+1}\sqrt{1-|\tau|^2\left(1-\frac{\Lambda_{pq}}{\Lambda_{p,q+1}}\right)}\quad,
\end{eqnarray}
where we have introduced the rescaled quantity $\tau:=\xi_1\xi_2/|J|^2 |t|^2$. The
operators (\ref{eq:way}) are correct up to terms of the order 
$O(|\tau|^2(1-|\tau|^2)(\xi_1-\xi_2)^2)$ which vanish if the tachyon is on-shell 
(see below) and are suppressed for large $p,q$.

\subsection[Topological $U(1)$ instanton charge and tachyon potential]{Topological nature of the $U(1)$ instanton charge and the tachyon potential} 

In \cite{Furuuchi:2000vc} it was shown that the instanton number (\ref{eq:top}), 
which is a surface term, 
 in noncommutative space also for the case of $U(1)$ 
gauge group has a topological origin.
The crucial input for this analysis is that the connections $\nabla_\alpha$ are 
of the form as given in our ansatz (\ref{eq:ans}). The tachyonic excitation does 
not contribute to the instanton number, or more exactly to the  Pontrjagin number 
for non-self-dual configurations, since it is located at the origin $z_1=z_2=0$.
The  shift operator(s) $S$  introduce an 
axial asymmetry in the sense that they either shift in the first or second copy of the 
Hilbert space corresponding to the two constituting noncommutative planes characterised
by $\theta_1$ and $\theta_2$. As mentioned above the choice of the shift operator 
(\ref{eq:shift}) is gauge dependent. In our case we have chosen 
the operator which shifts in the 
second Hilbert space. As a result the gauge field is located around $z_1=0$ 
and becomes a phase at the boundary of the plane parametrised by $z_2$. 
After integration over the $z_1$-plane the Pontrjagin charge  
reduces to a winding number of the gauge field around $S^1$ on the $z_2$-plane and is 
thus characterised by $\pi_1(U(1))$ (see \cite{Furuuchi:2000vc} for details). Although the
analysis in \cite{Furuuchi:2000vc} was done for instantons, 
i.e. self-dual solutions, all arguments
go through with our ansatz (\ref{eq:ans}) if the operators 
$Z_\alpha^\dagger$ become $\sim c_\alpha^\dagger$ far away from the centre. In terms
of the matrix elements $A_{pq}$ and $B_{pq}$  large distance to the centre means large
$p,q$. Thus the results of \cite{Furuuchi:2000vc} are valid for more general cases than 
self-dual configurations.

Based on the above arguments we can use the famous Bogomolnyi trick to write 
the potential as a sum of the Pontrjagin number and the anti-self-dual part of the
gauge field $F_{\mu\nu}$:
\begin{equation}
  \label{eq:V2}
  V=-\frac{1}{8g_{YM}^2}\int \textrm{Tr}\left[(F_{\mu\nu}-\tilde F_{\mu\nu})^2\right]
      +\frac{4\pi^2}{g_{YM}^2}\quad.
\end{equation}
The unstable and the final state fall in the class of solutions (\ref{eq:eomZ2}). If 
we assume that also for arbitrary $|t|^2$ the solution to (\ref{eq:eomZ}) falls into this 
class the tachyon potential (\ref{eq:V2}) simplifies to:
\begin{equation}
  \label{eq:V}
  V(|t|)=\frac{4\pi^2}{g_{YM}^2}\frac{\xi_1\xi_2}{2}
    \left[\left(\frac{|J|^2}{\xi_1\xi_2}-|t|^2\right)^2+\frac{2}{\xi_1\xi_2}\right]\quad.
\end{equation}
Regarding the dimension of the potential note that in five dimensions the Yang-Mills 
coupling is no longer dimensionless but has the dimension of a length. Clearly, the
extrema of the potential (\ref{eq:V}) are given by 
\begin{equation}
  \label{eq:ex}
  t=0\quad,\quad |t|^2=|\rho|^2=\frac{|J|^2}{\xi_1\xi_2}\quad,
\end{equation}
where the first one is the unstable extremum and the second 
one is a stable (local)  minimum. Note 
that in the latter case the tachyon mode (\ref{eq:tm}) coincides with the axial term
of the self-dual solution (\ref{eq:sol}). The local depth of the potential
is given by
{\setlength\arraycolsep{2pt}
\begin{eqnarray}
  \label{eq:dV}
  \Delta V&=&V(0)-V(|\rho|)\nonumber\\
  &=&\frac{2\pi^2}{g_{YM}^2}\ \frac{|J|^4}{\xi_1\xi_2}
  =\frac{2\pi^2}{g_{YM}^2}
  \left( \sqrt{\frac{\theta_1}{\theta_2}}-\sqrt{\frac{\theta_2}{\theta_1}}\right)^2\quad.
\end{eqnarray}}
Thus the depth of the potential exactly matches the mass defect $\Delta M$ (\ref{eq:dm})
of the $D0$-$D4$ system and is therefore in perfect agreement with Sen's conjecture 
\cite{Sen:1999mh}.

The approximate solution (\ref{eq:way}) we found above becomes exact for
the extrema (\ref{eq:ex}) and coincides at these points with the unstable state
(\ref{eq:dcomp}) or the final state (\ref{eq:sol}), respectively. When the tachyon is
off-shell (\ref{eq:way}) is valid only up to terms of order $O((\xi_1-\xi_2)^2)$, but
this error is suppressed for the  matrix elements where $q$ and $p$ are large. 
Anyway, (\ref{eq:way}) fulfils the requirements discussed above, such that it gives rise 
to a topological origin of the Pontrjagin number. Thus (\ref{eq:way}) 
smoothly parametrises a path in field space between the unstable and stable connection
(\ref{eq:dcomp}), (\ref{eq:sol}) such that for every $\tau\in[0,1]$  
(\ref{eq:way}) is a well defined $U(1)$ connection with constant 
integer Pontrjagin number one.

\subsection{Discussion}

The crucial input in our computation of the tachyon potential is that 
also for the $U(1)$ gauge theory the Pontrjagin number is 
a topological quantity. This also explains why the unstable state 
(\ref{eq:dcart}) which has the Pontrjagin number one (\ref{eq:top}) does not decay 
into the vacuum but to the BPS state (\ref{eq:sol}). In the $2+1$-dimensional case
as it was considered in \cite{Aganagic:2000mh} no such topological 
constraint exists. In contrary there 
exists a continuous path in field space, which connects the unstable solitonic state 
with the vacuum configuration, see also \cite{Gross:2000ss}.

Another crucial input is the ansatz (\ref{eq:ans}) which is based on 
symmetry arguments as given in \cite{Aganagic:2000mh}. So the 
relevant degrees of freedom in the condensation process are operators of the form 
(\ref{eq:Z}). According to our identification of the fluctuation spectrum with 
string excitations (\ref{eq:fl1}), (\ref{eq:scsp}) these are the massless $V_{44}$
strings. This is in contrast to the string field theory calculation of 
\cite{David:2000um}. 
In this case the zero level truncation approximation for a $(0,4)$ string field
also gives a quartic 
tachyon potential but the depth of the potential reproduces 
only $25$ percent of the mass defect (\ref{eq:dm}) 
in the considered range of the $B$-field. In \cite{David:2000um} 
it was supposed that the reason for this is the existence of a large number
of low-lying $(0,4)$-string states. But in our calculation the whole 
condensation effect comes from the massless $(4,4)$ strings. It would be nice
if a higher level string field theory calculation could resolve this seeming 
contradiction.

In \cite{Fujii:2001wp} also a quartic tachyon potential was obtained in a noncommutative 
field theory calculation. But the potential differs crucially from our 
result (\ref{eq:V}). It does not give the correct mass defect (\ref{eq:dm}). We believe 
that the 
reason for this is that in \cite{Fujii:2001wp} very strong conditions on the integrated out 
fields were imposed. It is doubtful whether a nontrivial solution to these conditions
exists. This would mean that the minimum of the tachyon potential as given in 
\cite{Fujii:2001wp} does not really describe a local minimum of the theory.

\section{Conclusions: A proposal and a puzzle}

We discussed several aspects of the unstable $D0$-$D4$ system with a  $B$-field 
background 
in terms of  noncommutative super Yang-Mills theory. 
In particular we studied zero-mode moduli and the 
fluctuation spectrum. For its stable limit (the small instanton),
i.e. when the  $B$-field becomes self-dual, we observed an equivalence of the
fluctuation spectrum with the one of the fluxon solution on 
$\mathbb{R}^2_\theta\times\mathbb{R}$ \cite{Gross:2000ph}. We argued 
that this equivalence of the 
spectra is due to $T$-duality but postponed a detailed discussion to future studies. 

Secondly we constructed a single $U(1)$ instanton for non-self-dual 
$B$-field backgrounds using the ADHM method. We were able to exactly compute the 
excitation  
spectrum for this state. It would be interesting if this could be compared with a string 
(field) theory calculation. 
The spectrum looks very different from the spectrum of the unstable and small instanton
 background,  not only quantitatively but also qualitatively. 
Although the classical solutions are continuously connected 
in  the limit of  self-dual $B$-field  the spectra are not. The number of states
significantly increases at the self-dual point compared to the excitation 
spectrum of the instanton for non-self-dual
$B$-field. Also additional zero-modes appear 
at the self-dual point. The scalar field  zero-modes 
associated with  translations of the $D0$-brane transverse to the $D4$-brane
are not present for the instanton with non-self-dual $B$-field. Thus the $D0$-brane
is confined in the $D4$-brane.

Finally we computed the tachyon potential for the unstable $D0$-$D4$ system. 
A crucial observation is that the Pontrjagin number in noncommutative 
$\mathbb{R}^4_\theta$ is topological  also for the $U(1)$ gauge theory  and  is thus
 conserved during the condensation process. Therefore the unstable $D0$-$D4$
condensates to a BPS state with same topological charge instead to the vacuum.
The in this way obtained quartic tachyon potential reproduces the whole mass defect 
of the formation of the $D0$-$D4$ bound state in the Seiberg-Witten limit.
The relevant degrees of freedom in the condensation process are the massless 
$(4,4)$ strings living on the $D4$-brane. This result is in contrast to string field 
calculations where only 25 percent of the mass defect could be reproduced and 
as the relevant degrees of freedom the low-lying $(0,4)$ strings were expected 
\cite{David:2000um}.

Based on the observed significant change in the number of excited states 
when the $B$-field becomes self-dual we propose the appearance of a phase transition
or at least a cross over at this point in the parameter space 
for the $B$-field background. This would accompany the deconfinement of
the $D0$-brane from the $D4$-brane. At the same time the effective theory on the
$D0$-brane changes from the Higgs branch to the Coulomb branch.
We hope to address this question in the future.

From a field theoretical point of view a puzzle remains. When considering one-loop 
instanton corrections a consistent regularisation plays an important role.
In \cite{Rebhan:2002uk,Rebhan:2002yw,Rebhan:2003bu,Rebhan:2004vn} 
such a susy-preserving dimensional regularisation for topological 
nontrivial backgrounds was developed by embedding the theory in a higher dimensional 
space with the same field content.\footnote{For an alternative regularisation suited 
for noncommutative theories see \cite{Behr:2005wp}.}  When using this regularisation, 
which also seems to apply  to the noncommutative case, only the
continuous spectrum contributes to quantum corrections. The contributions of
the discrete spectrum cancel each other because of the residual supersymmetry in the
BPS background. But if the continuous spectrum is gapless as we obtained it 
here in all cases it gives only scale-less integrals which vanish in dimensional 
regularisation. On the other hand, also in  noncommutative theories  
the bare parameters may renormalise non-trivially  
\cite{Grosse:2000yy,Bonora:2000ga,Sheikh-Jabbari:1999iw}. 
This especially means that in the 
$\ca{N}=2$ case the divergent renormalisation of the coupling constant would 
not be cancelled by an analogous contribution from the instanton determinants. 
The net result would be divergent. A possible resolution could be that noncommutativity
introduces a scale in the spectral density of the massless states. 
We have to postpone a detailed analysis of this issue  to future studies.

\section*{Acknowledgements}
I am grateful to L. Quevedo who was involved at intermediate stages of this work.
I especially thank M. Ihl, S. Uhlmann, K. Vogeler and M. Wolf for 
many helpful discussions.
I  thank O. Lechtenfeld, S. Petersen and A. Popov for useful comments on the script and
M. Rangamani for correspondence.

This work was done within the framework of the DFG priority program 
(SPP 1096) in string theory.

%\bibliographystyle{JHEP_rob}
%\bibliography{/home/itp/wimmer/science/bibtex/ego,/home/itp/wimmer/science/bibtex/ncinst,/home/itp/wimmer/science/bibtex/qft,/home/itp/wimmer/science/bibtex/books}

\providecommand{\href}[2]{#2}\begingroup\raggedright\endgroup

\end{document}